\documentclass[prd,nofootinbib,twocolumn]{revtex4}
\usepackage{amssymb}
\usepackage{natbib}

\usepackage[T1]{fontenc}

\def\vec#1{\ensuremath{\mathchoice{\mbox{\boldmath$\displaystyle#1$}}
{\mbox{\boldmath$\textstyle#1$}}
{\mbox{\boldmath$\scriptstyle#1$}}
{\mbox{\boldmath$\scriptscriptstyle#1$}}}}

\def\be{\begin{equation}}
\def\fe{\end{equation}}
\def\bea{\begin{eqnarray}}
\def\fea{\end{eqnarray}}

\def\O{{\mathcal{O}}}
\def\text#1{\textrm{#1}}
\def\mb{\mbox}

\def\X{{\!\mathit{X}}}
\def\Y{{\mathit{Y}}}

\def\A{{\mathrm{A}}}
\def\B{{\mathrm{B}}}

\def\Lagr{{\Lambda}}

\def\Hamil{\mathcal{H}}
\def\HamilH{{\Hamil_\H}}

\def\H{{\scriptscriptstyle{\mathrm{H}}}}
\def\LagrH{{\Lagr_\H}}

\def\Act{{\mathcal{I}}}

\def\T{\Theta}
\def\d{\delta}
\def\D{\Delta}
\def\qaq{{\;\;\textrm{and}\;\;}}
\def\E{{\mathcal{E}}}
\def\Et{{\tilde{\E}}}

\def\Dv{\Delta}
\def\entr{\alpha}
\def\entrn{\varepsilon}
\def\va{{\vec{a}}}
\def\vx{{\vec{x}}}
\def\vxi{ {\vec{\xi}} }

\def\dtau{{\tau}}
\def\dtauA{{\dtau_\X}}

\def\n{{\rm n}}
\def\p{{\rm p}}
\def\e{{\rm e}}
\def\c{{\rm c}}
\def\s{{\rm s}}

\def\equi{{\quad\Longleftrightarrow\quad}}

\def\eps{\epsilon}
\def\ext{{\mathrm{ext}}}

\def\vn{{\vec{n}}}
\def\vv{{\vec{v}}}
\def\vnabla{{{\nabla}}}
\def\vDv{{\vec{\Dv}}}

\def\vp{\vec{p}}
\def\po{p_0}

\def\Rate{\mathit{\Gamma}}
\def\csum{\sum}
\def\Press{\Psi}

\def\S{{\mathrm{S}}}
\def\N{{\mathrm{N}}}

\def\cf{f}
\def\vcf{\vec{\cf}}

\def\vf{\vec{f}}
\def\vfH{\vf_\H}

\def\vef{\widehat{\vf}}
\def\vefH{\vef_\H}
\def\eg{\widehat{g}}

\def\FF{{\mathfrak{F}}}
\def\mutual{{\mathrm{mut}}}

\def\mut{\widetilde{\mu}}

\def\st{{\widetilde{s}}}

\def\vV{{\vec{V}}}

\def\vort{w}
\def\Vort{W}

\def\Circ{{\mathcal{C}}}

\def\Lie{\pounds}
\def\form{\underline}

\def\Hel{{\mathcal{H}}}

\def\vJ{{\vec{J}}}
\def\vQ{{\vec{Q}}}

\def\vq{{\vec{q}}}

\def\enth{{\mathrm{w}}}

\def\Z{{\mathcal{Z}}}
\def\z{z}

\def\V{\nu}
\def\prho{\varrho}
\def\pDv{{\vec{w}}}
\def\sff{{(0)}}

\def\1{{(1)}}
\def\2{{(2)}}

\def\cov{{\mathrm{cov}}}
\def\ren{{\mathrm{ren}}}

\def\R{{\!\!R}}
\def\Jac{\mathcal{J}}

\begin{document}

\title{Variational description of multi-fluid hydrodynamics: Uncharged fluids.}

\author{Reinhard Prix}

\affiliation{Department of Mathematics, University of Southampton, SO17
  1BJ, UK}

\date{Nov. 26, 2003}
\email{Reinhard.Prix@aei.mpg.de}

\begin{abstract}
We present a formalism for Newtonian multi-fluid hydrodynamics
derived from an \emph{unconstrained} variational principle.
This approach provides a natural way of obtaining the general equations
of  motion for a wide range of hydrodynamic systems containing an
arbitrary number of interacting fluids and superfluids. 
In addition to spatial variations we use ``time shifts'' in the
variational principle, which allows us to describe dissipative
processes with entropy creation, such as chemical reactions, 
friction or the effects of external non-conservative forces. 
The resulting framework incorporates the generalization of the
\emph{entrainment} effect originally discussed in the case of the
mixture of two superfluids by Andreev and Bashkin. 
In addition to the conservation of energy and momentum, we derive the
generalized conservation laws of vorticity and helicity, and the
special case of Ertel's theorem for the single perfect fluid. 

We explicitly discuss the application of this framework to thermally
conducting fluids, superfluids, and superfluid neutron star
matter. The equations governing thermally conducting fluids are found 
to be more general than the standard description, as the effect of
entrainment usually seems to be overlooked in this context. 
In the case of superfluid $^4$He we recover the Landau--Khalatnikov
equations of the two-fluid model via a translation to the
``orthodox'' framework of superfluidity, which is based on
a rather awkward choice of variables. Our two-fluid model for
superfluid neutron star matter allows for dissipation via mutual 
friction and also ``transfusion'' via $\beta$-reactions between the
neutron fluid and the proton-electron fluid.
\end{abstract}


% % PACS codes here, in the form: \PACS code \sep code
% \end{keyword}

\maketitle

\section{Introduction}

The main purpose of this work is to develop a formalism that allows one
to derive the equations of motion for a general class of
multi-constituent systems of interacting charged and uncharged fluids,
such as conducting and non-conducting fluids, multi-fluid plasmas,
superfluids and superconductors. For the sake of clarity of
presentation we restrict ourselves here to uncharged fluids, while the
case of charged fluids and their coupling to the electromagnetic field
will be treated in a subsequent paper \cite{prix03:_variat_II}. 

Long after the completion of classical Hamiltonian particle mechanics,
the quest of finding a variational (or ``Hamlitonian'') description of
hydrodynamics has surprisingly been a long-standing problem, which
started only a few decades ago to be fully understood. 
The reason for this can be traced to the nature of the hydrodynamic
equations, which are most commonly expressed in their Eulerian form in
terms of the \emph{density} $\rho$ and \emph{velocity} $\vv$, where
the information about the underlying flowlines has been hidden. Fluid
particle trajectories, i.e. flowlines, can still be recovered by
integrating the velocity field, but they are not independent
quantities of the Eulerian description. However, it turns out that the
``true'' fundamental field variables of Hamiltonian hydrodynamics are
the flowlines, which determine $\rho$ and $\vv$ as derived quantities.

Consider as an example the Lagrangian density $\Lagr$ describing a
barotropic perfect fluid, which in analogy to classical mechanics one 
would postulate to be
$$
\Lagr(\rho,\vv) = {1\over 2}\rho \vv^2 - \E(\rho)\,,
$$
where $\E(\rho)$ represents the internal energy density of the fluid.
We note that the internal energy defines the chemical potential $\mut$
and the pressure $P$ as
$$
d \E = \mut \, d \rho\,,\qaq P + \E = \rho \, \mut\,.
$$
The corresponding action is defined in the usual way as
\mb{$\Act\equiv\int \Lagr\,d V\,d t$}, and the variation $\d\Lagr$ of
the Lagrangian density is
$$\d\Lagr = \rho\vv\cdot \d\vv + (\vv^2/2 -\mut)\, \d\rho\,.$$ 
Requiring the action $\Act$ to be stationary with respect to
\emph{free variations} $\d\rho$ and $\d\vv$ is immediately seen to
be useless, as this leads to the over-constrained equations of motion
\mb{$\rho\vv=0$} and \mb{$\mut=\vv^2/2$}. 
In fact, it has been shown \cite{schutz77:_variat_aspec} that 
an unconstrained variational principle with $\rho$ and $\vv$ as
the fundamental variables cannot produce the Eulerian hydrodynamic
equations.  
The reason for this is rather intuitive, as it is evident that free
variations of density and velocity  probe configurations with
different masses (i.e. different numbers of particles),
which is not an actual degree of freedom of the dynamics of the
system. Therefore the variational principle has to be constrained or
reformulated in some way in order to restrict the variations to the
physically meaningful degrees of freedom. 

The historic approach to this problem in Newtonian physics has been to
supplement the Lagrangian with appropriate constraints using Lagrange 
multipliers. This method was pioneered by Zilsel
\cite{zilsel50:_liquid_helium_ii} in the context of the two-fluid
model for superfluid $^4$He, who used the constraints of conserved
particles (i.e. mass) and entropy.  However, as pointed out by Lin
\cite{lin63:_hydrod_helium_ii}, this is generally insufficient,
as it results in equations of motion restricted to 
\emph{irrotational flow} in the case of uniform entropy. Lin showed  
that one has to add a further constraint, namely the ``conservation
of identity'' of fluid particles in order to obtain the most general
hydrodynamic equations. We can label particles by their initial
positions $\va$, and so we can write their flowlines as
\mb{$\vx=\vx(\va,t)$}. The famous ``Lin constraint'' 
is \mb{$\partial_t \va + \vv\cdot \vnabla \va = 0$},
i.e. the identity or label of a particle is conserved under its
transport.  For reviews of this approach and its relation to the
``Clebsch representation'' we refer the reader to 
\cite{seliger68:_variat,salmon88:_hamil_fluid_mechan,zakharov97:_hamil},
and references therein.

Although this method produces the correct equations of motion, it does
not seem very natural due to the rather ad~hoc introduction of
constraints, and the need for unphysical auxiliary fields (the Lagrange
multipliers).
It was pointed out by Herivel~\cite{herivel55:_deriv_equat}
that the \emph{Lagrangian} as opposed to Eulerian formulation of
hydrodynamics results in a much more natural variational description,
and this approach was further developed and clarified by Seliger and
Whitham \cite{seliger68:_variat}.  
Instead of using $\rho$ and $\vv$ as fundamental variables,
hydrodynamics can also be understood as a field theory in terms of the
\emph{flowlines} $\vx(\va,t)$, or equivalently
\mb{$\va=\va(\vx,t)$}. It turns out that this formulation allows for a
perfectly natural \emph{unconstrained} variational principle. 
This seems rather intuitive considering that hydrodynamics is a
smooth-averaged description of a many-particle system, which is
described by a variational principle based on the particle
trajectories, i.e. $\vx_N$ and $\dot{\vx}_N$. 

We can express the velocity and density in terms of the flowlines as
\mb{$\vv =\partial_t \vx(\va,t)$} and 
\mb{$\rho(\vx,t) = \rho_0(\va) / \det({\Jac^i}_j)$}, where 
\mb{${\Jac^i}_j = \partial x^i / \partial a^j$} is the Jacobian matrix
corresponding to the map \mb{$\va\mapsto\vx(\va,t)$} between the
physical space $\vx$ and the ``material space'' $\va$. Any further
comoving quantities like the entropy $s$ are determined in terms of
their initial value $s_0(\va)$. Substituting these expressions into
the Lagrangian $\Lagr$, one obtains an unconstrained variational
principle for the field $\vx(\va,t)$, which results in the correct
equations of motion. 
It is interesting to note that this approach implicitly satisfies Lin's
constraint, as we are varying the particle trajectories
$\vx(\va,t)$, along which $\va$ is a constant by construction. Also,
we do not need to impose an a~priori constraint on the conservation
of mass, as it is automatically satisfied by these ``convective''
variations: shifting around flowlines obviously conserves the number
of flowlines, and therefore the number of particles. 
One can actually \emph{derive} the Lin constraint by transforming this
Lagrangian framework back into a purely Eulerian variational principle
\cite{seliger68:_variat,salmon88:_hamil_fluid_mechan}, which
shows that these two approaches are formally equivalent. 

As pointed out by Bretherton \cite{bretherton70:_hamilton_fluid}, one
can even more conveniently use a ``hybrid'' approach, in which the
Lagrangian is expressed in terms of the Eulerian hydrodynamic
quantities $\vv$, $\rho$, $s$ etc, but one consider them as functions
of the underlying flowlines. Their variations are therefore naturally
\emph{induced} by variations $\vxi$ of the flowlines $\vx(\va,t)$. 
In general relativity the same idea was pioneered by Taub
\cite{taub54:_gr_variat_princ}, and has subsequently been largely 
developed and extended by Carter  
\cite{carter73:_elast_pertur,carter83:_in_random_walk,carter89:_covar_theor_conduc},
who also coined the term ``convective variational principle'' for this
approach. Carter and Khalatnikov
\cite{carter92:_equiv_convective_potential} have further demonstrated the
formal equivalence of the convective approach and the more common
Clebsh formulation that results from an Eulerian variational approach.
A ``translation'' of the covariant convective formalism into a
Newtonian framework  (albeit using a spacetime-covariant language
close to general relativity) is also available
\cite{carter94:_canon_formul_newton_superfl,carter04:_covar_newtonI}.
The convective approach in relativity has independently been developed 
by Kijowski \cite{kijowski79:_sympl_framew}, and Hamiltonian
formulations have been constructed by Comer and Langlois 
\cite{comer93:_hamil_multi_constituent} and Brown
\cite{brown93:_action_fluids}.  
Here we are using the convective (or ``hybrid'') variational
principle in order to derive the Newtonian multi-fluid equations, 
and our notation and formalism follows most closely the framework
developed by Carter. 

We conclude our example of the simple barotropic fluid by using the
convective variational principle to derive the Euler equation. 
The expressions for (Eulerian) variations of density and velocity
\emph{induced} by infinitesimal spatial displacements $\vxi$ of the
flowlines are well known\footnote{A generalization of these
  expressions to include time-shifts is derived in  
Appendix~\ref{sec:Variations}} (e.g. see \cite{friedman78:_lagran}),
namely
$$
\d\rho = - \vnabla\left( \rho \vxi \right) \,,\qaq
\d \vv = \partial_t \,\vxi + (\vv\cdot\vnabla) \vxi - (\vxi\cdot\vnabla) \vv\,.
$$
Inserting these expressions into the variation of the action
\mb{$\d\Act=\int\d\Lagr\,d V\,d t$} with $\d\Lagr$ given above, and
after some integrations by parts and dropping total divergences and
time derivatives (which vanish due to the boundary conditions), we find  
\begin{eqnarray*}
\d\Act =& & -\int \vxi\cdot\biggl[\rho(\partial_t + \vv\cdot\vnabla)\vv
  + \rho\vnabla\mut \\
&& + \vv \left\{\partial_t \rho +
    \vnabla\cdot(\rho\vv) \right\}\biggl]\,d V\,d t\,.
\end{eqnarray*}
If we assume conservation of mass\footnote{This will be seen to be a
  consequence of the variational principle rather than an a-priori
  assumption when time-shift variations are included.}, i.e. 
\mb{$\partial_t \rho + \vnabla\cdot(\rho\vv)=0$}, then stationarity of  
the action (i.e. \mb{$\d\Act=0$}) under free variations $\vxi$
directly leads to Euler's equation, namely
$$
(\partial_t + \vv\cdot\vnabla)\,\vv + {1\over\rho}\vnabla P = 0\,,
$$
where we have used the thermodynamic identity 
\mb{$\rho\vnabla\mut = \vnabla P$}. This shows that an unconstrained
convective variational principle produces to the correct hydrodynamic
equations of motion in a surprisingly simple and straightforward way.

The spatial variations $\vxi$ have three degrees of freedom, resulting
in one vector equation, which represents the conservation of
momentum. In order to complete the description we will need a fourth
variational degree of freedom to produce the missing energy
equation. This can be achieved by considering time-shifts,
which are a natural part of the covariant relativistic approach, but
which we have to be considered explicitly in the conventional ``3+1''
language of Newtonian space-time. 
These time-shifts variations allow us to take this formalism to its
full  generality, as we can now describe even dissipative processes
with entropy creation, particle transformations (i.e. chemical
reactions), resistive frictional forces etc. These dissipative systems
are of course still \emph{conservative} as long as one includes
entropy, which is why they can be described by an action
principle. The second law of thermodynamics, however, is obviously not
contained in the action principle and has to be imposed as an
additional equation on the level of the equations of motion.  

We note that the equations we derive here do not explicitly
include shear- and bulk-viscosity effects. However, the current
\emph{form} of the equations is in principle general enough to allow 
for both of these effects: bulk viscosity is caused by heat flow or
chemical reactions due to thermal or chemical disequilibrium, both of
which can already be described in the current formulation. Shear
viscosity on the other hand has to be introduced as an ``external''
force, the problem therefore consists in prescribing a  physically
reasonable model for a multi-fluid generalization of the shear
stresses. Including viscosity should therefore not be a matter 
of actually \emph{extending} the current framework but rather of
appropriately applying it in order to describe such processes. An
explicit discussion of this is postponed to future work. 
Further work is also necessary in order to extend this formalism to
include elasticity (as pioneered in the relativistic framework
\cite{carter72:_found_elasticity}), and especially to allow for an
elastic medium interpenetrated by fluids as encountered in the inner
neutron star crust, or any type of conducting solid.
As shown in \cite{carter95:_kalb_ramond}, a Kalb-Ramond type
extension is required for the macroscopic treatment of quantized
vortices in superfluids. With the present formalism we can describe
superfluids either on the local irrotational level, or on the
smooth-averaged macroscopic level by neglecting the (generally small)
anisotropy induced by the quantized vortices.  

The plan of this paper is as follows: in
Sect.~\ref{sec:GeneralMultiConst} we derive the general form of the
equations of motion for multi-constituent systems using the convective
variational principle. In Sect.~\ref{sec:TotalConservationLaws} we show the
conservation of energy and momentum implied by these equations.
In Sect.~\ref{sec:FlowlineConservations} we derive conserved
quantities under transport by the flow, namely the vorticity and
helicity. We then give the explicit functional form of the Lagrangian
density for hydrodynamic systems in Sect.~\ref{sec:Hydrodynamics}, and
in Sect.~\ref{sec:Applications} we discuss several applications of the
foregoing formalism to particular physical systems.

\section{Variational description of multi-constituent systems}
\label{sec:GeneralMultiConst}

\subsection{Kinematics}
\label{sec:Kinematics}

We want to describe systems consisting of several constituents
distinguished by suitably chosen labels, and we use capital letters
$\X,\Y,...$ as indices which run over these constituents labels. 
As the fundamental quantities of the kinematic description we choose
the constituent densities $n_\X$ and the associated transport currents
$\vn_\X$,  which are related to the respective velocities $\vv_\X$ as
\be
\vn_\X = n_\X \vv_\X\,,\quad\text{where}\quad \X \in
\{\textrm{constituent labels}\}\,. 
\label{equ0}
\fe
Not all constituents can necessarily move independently from each other,
i.e. not all velocities $\vv_\X$ have to be different: viscosity and
friction due to particle collisions on the microscopic level can
effectively bind constituents together on very short
timescales. We therefore distinguish between the notions of
\emph{constituents} $\X$, characterizing classes of microscopic
particles, and \emph{fluids}, which are sets of constituents with a
common velocity.  

We note that in this framework entropy can be described very naturally
as a constituent for which we reserve the label $\X=\s$, and we write 
\be
n_\s = s\,,
\fe
where $s$ is the entropy density. In this context it is instructive to
think of the entropy as a gas of particle-like thermal
excitations (e.g. phonons, rotons etc.), which makes its description
as a constituent on the same footing with particle number densities
quite intuitive.

\subsection{Dynamics}

The dynamics of the system is governed by an action $\Act$ defined as
\begin{equation}
  \label{eq:DefAct}
  \Act = \int \LagrH \, d V\, d t\,,
\end{equation}
in terms of the hydrodynamic Lagrangian $\LagrH$.
The Lagrangian density $\LagrH$ depends on the kinematic variables,
which are the densities $n_\X$ and the currents $\vn_\X$,
i.e. \mb{$\LagrH=\LagrH(n_\X, \vn_\X)$}. The total differential of
$\LagrH$ defines the  \emph{dynamical} quantities $p^\X_0$
(``energy'') and $\vp^\X$ (``momentum'') per fluid particle as the
canonically conjugate variables to $n_\X$ and $\vn_\X$, namely
\be
d\LagrH\!\!=\!\csum\!\left(\po^\X\!dn_\X\!+\!\vp^\X\!\!\cdot d \vn_\X
\right)\!,
\;\;\textrm{so}\;\;
\po^\X\!\!=\!{\partial \LagrH \over \partial n_\X},\;\;
\vp^\X\!\!=\!{\partial \LagrH \over \partial \vn_\X},
\label{equVarL}
\fe
where here and in the following the sum over repeated constituent
indices is explicitly indicated by a $\Sigma$, i.e. no automatic
summation convention applies to constituent indices. 

\subsection{The convective variational principle}

As we have seen in the introduction, one cannot apply the
standard variational principle to $\LagrH$ in terms of the Eulerian
hydrodynamics variables $n_\X$ and $\vn_\X$. From (\ref{equVarL}) it is
obvious that allowing {\em free} variations of densities $\d n_\X$ and
currents $\d \vn_\X$  would lead to the trivial equations of motion
\mb{$\po^\X=0$} and \mb{$\vp^\X=0$}. 
Instead, we consider the Lagrangian to be a functional of the
underlying \emph{flowlines} \mb{$\vx^\X=\vx^\X(\va^\X,t)$}, and 
therefore admit only variations $\d n_\X$, $\d \vn_\X$
that are {\em induced} by infinitesimal displacements of the
flowlines. These ``convective''  variations naturally conserve the
number of particles (i.e. the number of flowlines) and no constraints
are required in the variational principle as was discussed in more
detail in the introduction. 

We apply infinitesimal spatial displacements $\vxi_\X$ and time-shifts
$\dtau_\X$ to the flowlines of the constituent $\X$. The 
resulting induced variations of density and current have been derived
in Appendix~\ref{sec:Variations}, namely the density variation
(\ref{eq:dDensity}) for constituent $\X$ is
\be
\d n_\X = - \vnabla\cdot\left[n_\X\vxi_\X \right] 
+\left[ \vn_\X\cdot\vnabla\dtauA - \dtauA\partial_t n_\X\right ]\,,
\fe
while the current variation $\d\vn_\X$ is given by (\ref{eq:dCurrent})
and reads as 
\bea
\d\vn_\X &=& n_\X\partial_t \vxi_\X + (\vn_\X\cdot\vnabla)\, \vxi_\X
- (\vxi_\X\cdot\vnabla)\vn_\X \nonumber\\
&& - \vn_\X (\vnabla\cdot\vxi_\X) -
\partial_t \left( \vn_\X \dtauA\right)\,.
\fea
Inserting these expressions into the variation of the Lagrangian
(\ref{equVarL}) and integrating by parts, we can rewrite the induced
variation $\d\LagrH$ in the form 
\be
\d\LagrH = \csum \left(  g^\X \,  \dtauA - \vcf^\X\cdot\vxi_\X \right) + 
\partial_t R + \vnabla\cdot\vec{R}\,.
\label{equdL}
\fe
The time derivative and divergence terms will vanish in the action
integration (\ref{eq:DefAct}) by the appropriate boundary conditions
(i.e. $\vxi=0$ and $\dtau=0$) and are irrelevant as far as the
variational principle is concerned, but for completeness we note that
their explicit expressions are 
\bea
R &\equiv& \csum\left( n_\X \vp^\X\cdot \vxi_\X - \vn_\X\cdot\vp^\X \dtauA \right)\,,\\
\vec{R} &\equiv& \csum [
\vn_\X \,(\po^\X\, \dtauA + \vp^\X\cdot\vxi_\X)  \nonumber\\
 && - \vxi_\X (n_\X \po^\X + \vn_\X\cdot\vp^\X) ]\,.
\fea
The induced action variation therefore has the form
\be
\d\Act = \csum \int \left(  g^\X\, \dtau_\X  - \vcf^\X\cdot\vxi_\X
\right)\,d V\,d t\,, \label{equActionVar} 
\fe
where the force densities $\vcf^\X$ (acting {\em on} the constituent)
and the energy transfer rates $g^\X$ ({\em into} the constituent) are
found explicitly as  
\bea
\vf^\X\!\! &=&\! n_\X\!\left(\partial_t \vp^\X\!-\!\vnabla \po^\X \right)\!-
\vn_\X\!\times(\vnabla\!\times\vp^\X)\!+\!\vp^\X\!\Rate_\X,
\label{equfA}\\
g^\X\!\!&=&\! \vv_\X\!\cdot\left(\vf^\X - \vp^\X \Rate_\X \right) -
\po^\X \Rate_\X\,,
\label{equgA}
\fea
where $\Rate_\X$ is the particle creation rate for the constituent
$\X$, i.e. 
\be
\Rate_\X \equiv  \partial_t n_\X + \vnabla\cdot \vn_\X\,. \label{eq:DefRateX}
\fe
The force density $\vf^\X$ is the total momentum change
rate of the constituent $\X$, and we see that the last term in 
(\ref{equfA}), i.e. the ``rocket term'' $\vp^\X \Rate_\X$, represents
a contribution that is purely due to the change of the particle number. 
Therefore it will be convenient to define the purely ``hydrodynamic
force'' $\vfH^\X$, as
\be
\vfH^\X \equiv n_\X \left(\partial_t \vp^\X - \vnabla \po^\X \right) -
\vn_\X \times (\vnabla\times\vp^\X)\,.
\label{eq:DeffH}
\fe
With this definition we can now write the force density (\ref{equfA})
and energy transfer rate (\ref{equgA}) in the form 
\bea
\vf^\X &=& \vfH^\X + \vp^\X \,\Rate_\X\,, \label{eq:fX}\\
g^\X &=& \vv_\X\cdot \vfH^\X -  \po^\X\, \Rate_\X\,.  \label{eq:EnFlux}
\fea

\subsection{The equations of motion}

Up to this point we have developed only purely mathematical identities
without a specific physical content. The equations of motion 
are obtained by imposing which type of invariance the action $\Act$
should satisfy under certain infinitesimal variations. 
The most general equations are obtained by requiring that a 
{\em common} displacement $\vxi_\X=\vxi$ and time shift
$\dtau_\X=\dtau$ of all constituents should result in an action 
variation of the form
\be
\d\Act = \int \left(  g_\ext\, \dtau - \vcf_\ext\cdot\vxi \right)\, d V\,
d t\,,
\fe
where $\vcf_\ext$ and $g_\ext$ are interpretable as the external
force density and energy transfer rate. This generalizes the more
common action principle of \emph{isolated} systems, in which the
external influences $\vcf_\ext$ and $g_\ext$ vanish and therefore the
equations of motion are obtained by requiring the action to be
\emph{invariant} under small variations.
``External'' here is meant in the sense of not being included in the
total Lagrangian, which could also mean, for example viscous or
gravitational forces. 
The resulting minimal equations of motion obtained from
comparing with (\ref{equActionVar}) are therefore found as
\be
\csum \vcf^\X = \vcf_\ext\,,\qaq
\csum g^\X = g_\ext\,. \label{eq:GeneralEOM}
\fe
Together with (\ref{equfA}) and (\ref{equgA}) this represents the
Euler-Lagrange equations associated with this variational principle.
If all constituents $\X$ form a single fluid, namely all constituents
have a common velocity, then only common displacements of all
constituents make sense in the variational principle. For this class of 
\emph{non--conducting} models, (\ref{eq:GeneralEOM}) represent 
the full equations of motion obtainable from the variational
principle. In order to complete the model, one has to specify the
hydrodynamic Lagrangian $\LagrH$, the external interactions
$\vcf_\ext$ and $g_\ext$, and the creation rates $\Rate_\X$ as
functions of the kinematic variables.

In the case of \emph{conducting} models, at least some of the 
constituents are allowed to move independently, the system therefore
consists of more than one fluid. This increases correspondingly the 
number of degrees of freedom, and more equations of motion are
required. They are obtained very naturally from the variational
principle, as independent displacements (in space and time) are
permitted for \emph{each fluid}. Therefore the resulting force
acting on each fluid can be prescribed by the model, subject to the
restriction only of satisfying the minimal equations of motion
(\ref{eq:GeneralEOM}). 

As an example, consider the case of a simple conducting model
consisting of two fluids, where we use $\X$ and $\Y$ are constituent
indices running only over the respective constituent labels, i.e. 
\mb{$\X \in \{\textrm{fluid 1}\}$} and 
\mb{$\Y \in \{\textrm{fluid 2}\}$}. We then have the respective force
densities acting on each of the two fluids as 
\mb{$\vcf_\1 = \csum_\X \vcf^\X$} and \mb{$\vcf_\2 = \csum_\Y \vcf^\Y$},
which by (\ref{eq:GeneralEOM}) have to satisfy
$\vcf_\1 + \vcf_\2 = \vcf_\ext$. Therefore there are now exactly two
force densities (e.g. $\vcf_\1$ and $\vcf_\ext$) freely specifiable in
the model, corresponding to the additional degrees of freedom of two
fluids. In this case $\vcf_\1$ could for example represent a mutual
force the two fluids exert on each other, e.g. a resistive friction
force. 

\section{Conservation of energy and momentum}
\label{sec:TotalConservationLaws}

Using the explicit expression (\ref{equfA}) for the force density $\vcf^\X$, we
can write
\bea
\csum \vcf^\X &=& \partial_t\left( \csum n_\X\,\vp^\X \right) 
+ \nabla_j \left(\csum n_\X^j\, \vp^\X \right) \nonumber\\
& & - \csum \left(n_\X \vnabla \po^\X + n_\X^j \vnabla p^\X_j
\right)\,. \label{eq:34}
\fea
We define the ``generalized pressure'' $\Press$
via the total Legendre transformation of $\LagrH$, namely
\be
\Press \equiv \LagrH - \csum \left( n_\X \po^\X + \vn_\X\cdot \vp^\X\right)\,,
\label{equDefPress}
\fe
which is seen from (\ref{equVarL}) to result in the total differential 
\begin{equation}
  \label{eq:dPsi}
  d\Press = - \csum \left( n_\X d p^\X_0 + \vn_\X\cdot d \vp^\X \right)\,,
\end{equation}
and therefore the last sum in (\ref{eq:34}) is simply $\vnabla \Press$.
We can now cast the force equation (\ref{eq:GeneralEOM}) in the form
of a conservation law for the total momentum, namely  
\be
\partial_t J_\H^i + \nabla_j T_\H^{i j} = f_\ext^i\,,
\label{eq:MomCons}
\fe
where the hydrodynamic momentum density $\vJ_\H$ and stress tensor
$T_\H^{i j}$ are defined as 
\begin{equation}
  \label{eq:Tmunu}
  \vJ_\H \equiv \csum n_\X \vp^\X\,,\qaq
T_\H^{i j} \equiv \csum n_\X^i p^{\X\,j} + \Press\, g^{i j}\,,
\end{equation}
and where $g_{i j}$ are the components of the metric tensor determining
the relation between physical distance $d l$ and coordinate intervals
$d x^i$, i.e. \mbox{$d l^2 = g_{i j}\, d x^i \, d x^j$}. In Cartesian
coordinates this is simply $g_{i j}=\delta_{i j}$. 
A proof of the symmetry of the stress tensor $T_H^{i j}$ together with
a more elegant derivation of momentum conservation as a Noether
identity of the variational principle is given in
Appendix~\ref{sec:VariationalTij}.

Using expressions (\ref{equfA}) and (\ref{equgA}), we can further show
that  
\begin{eqnarray}
\csum g^X &=&
\csum \left[ \vn_\X\cdot \partial_t \vp^\X - \vn_\X\cdot\vnabla p^\X_0 -
  \Rate_\X p^\X_0  \right] \nonumber \\
&=& \left( \partial_t \csum \vn_\X\cdot\vp^\X\right) 
- \vnabla\cdot\left(\csum \vn_\X p^\X_0\right) \nonumber \\
&& - \csum ( p^\X_0 \partial_t n_\X + \vp^\X\cdot\partial_t \vn_\X )\,,
\end{eqnarray}
and we see from (\ref{equVarL}) that the last sum simply represents
$\partial_t \LagrH$. We can therefore rewrite the energy equation 
(\ref{eq:GeneralEOM}) in the form of a conservation law, namely
\begin{equation}
  \label{eq:EnergyCons2}
  \partial_t E_\H + \vnabla\cdot\vQ_\H = g_\ext\,,
\end{equation}
where the hydrodynamic energy density $E_\H$ and energy flux
$\vQ_\H$ are given by 
\begin{equation}
  \label{eq:T0mu}
  E_\H\!=\!\csum \vn_\X\!\cdot\vp^\X - \LagrH\,,\;\;\textrm{and}\;\;
  \vQ_\H\!=\!\csum (-p^\X_0) \vn_\X\,.
\end{equation}
We see that the energy density $E_\H$ has quite naturally the form of
a  hamiltonian, i.e. \mb{$\HamilH(n_\X, \vp^\X) = E_\H$}, as it is the 
Legendre-transformed (with respect to the momenta) of the Lagrangian
$\LagrH$.  

\section{Conservation along flowlines}
\label{sec:FlowlineConservations}

In addition to the total energy-momentum conservation,
derived in the previous section, we can find further conserved
quantities for individual constituents, for which conservation holds  
under transport by the fluid flow. Because the following derivations
apply to individual constituents instead of the sum over all
constituents, we will omit the constituent index $\X$ in this section 
in order to simplify the notation.   

Transport of a quantity by the fluid flow is closely related to the
Lie derivative with respect to the fluid velocity, therefore these 
conservation laws are most easily derived using the language and
theorems of differential forms instead of vectors. We will use this
formalism in deriving the transport-conservation laws, but we also
give the essential steps and results translated in the more common
vector- and index-notation, so that familiarity with exterior calculus
should not be necessary (albeit helpful) for reading this section. 

\subsection{Kelvin-Helmholtz vorticity conservation}
\label{sec:Vorticity}

We define the vorticity 2-form $\form{\vort}$ (with components
$\vort_{i j}$) as the exterior derivative (denoted by $d$) 
of the momentum 1-form $\form{p}$ (with components $p_i$), namely 
\begin{equation}
  \label{eq:DefVort}
  \form{\vort} \equiv d \form{p}\,,\quad\textrm{i.e.}\quad
 \vort_{i j} \equiv 2 \nabla_{[i} p_{j]}\,, 
\end{equation}
where $[i j]$ denotes antisymmetric averaging, i.e. 
\mb{$2 A_{[i}B_{j]}=A_i B_j - A_j B_i$}.
In three dimensions we can define the more common vorticity
\emph{vector} $\vec{\Vort}$ as the \emph{dual} (with respect to the
volume form $\eps_{i j k}$) of the 2-form $\form{\vort}$, namely
\begin{equation}
  \label{eq:DefDualVortVector}
 \Vort^i \equiv {1\over2}\eps^{i j k} \vort_{j k} = \left( \vnabla
  \times \vp\right)^i\,.
\end{equation}
The volume form is defined as 
\be
\eps_{i j k} = \sqrt{g}\,[i,j,k]\,,
\label{eq:DefVolumeForm}
\fe
where $g = \det(g_{i j})$ and $[i,j,k]$ is the sign of the permutation
of $\{1,2,3\}$, which is zero if two indices are equal. 
The duality between $\form{\vort}$ and $\vec{\Vort}$ implies 
\begin{equation}
  \vort_{i j} = \eps_{i j k} \Vort^k\,,
  \label{eq:DualVort}
\end{equation}
which is easily verified by inserting (\ref{eq:DefDualVortVector}).
We note that due to the Poincar\'e property (namely \mb{$d d = 0$}), the
exterior derivative of the vorticity 2-form vanishes identically, i.e.
\begin{equation}
  d\form{\vort} = 0 \quad\Longleftrightarrow \vnabla\cdot\vec{\Vort} = 0\,.
\end{equation}
We can rewrite the expression (\ref{eq:DeffH}) for the
hydrodynamic force $\vfH$ in  the language of forms as
\begin{equation}
  \partial_t \form{p} + \vv\rfloor d\form{p} - d \po = {1\over n}\form{f}_\H\,,
  \label{eq:forceq1}
\end{equation}
where $\rfloor$ indicates summation over adjacent vector- and form-
indices, i.e. in this case 
\mb{$(\vv\rfloor d\form{p})_i = 2 v^j \nabla_{[j}p_{i]}$}.
In the following it will be convenient to separate the force per
particle into its non-conservative part $\form{\FF}$ and a 
conservative contribution $d\phi$, namely
\begin{equation}
  {1\over n}\form{f}_\H = d\phi + \form{\FF}\,.
  \label{eq:ForceSeparation}
\end{equation}
The Cartan formula for the Lie derivative of a $p$-form applied to the
1-form $\form{p}$ yields 
\begin{equation}
  \Lie_\vv\, \form{p} = \vv \rfloor d\form{p} + d(\vv \rfloor \form{p})\,,
\end{equation}
which in explicit index notation reads as 
\mb{$\Lie_\vv p_i = 2 v^j \nabla_{[j} p_{i]} + \nabla_i ( v^j p_j)$}.
Using this identity and (\ref{eq:ForceSeparation}) we rewrite the
force equation (\ref{eq:forceq1}) more conveniently as 
\begin{equation}
  (\partial_t + \Lie_\vv)\, \form{p} = d Q + \form{\FF}\,,
  \label{eq:Helmholtz1}
\end{equation}
where the scalar $Q$ is given by $Q = \po +   \vv\rfloor\form{p}$.
Lie derivatives and partial time derivatives commute with exterior
derivatives, so we can apply an exterior derivative to
(\ref{eq:Helmholtz1}) and obtain  the Helmholtz equation of vorticity
transport, namely   
\begin{equation}
  \label{eq:HelmholtzIa}
  (\partial_t + \Lie_\vv) \,\form{\vort} = d \form{\FF}\,,
\end{equation}
which shows that the vorticity is conserved under transport by the
fluid if and only if the hydrodynamic force per particle acting on the
fluid is purely conservative, i.e. if \mb{$\form{\FF} = 0$}.
In its more common dual form, this equation can be written as
\begin{equation}
  \label{eq:HelmholtzIb}
  \partial_t \vec{\Vort} - \vnabla\times\left(\vv \times
  \vec{\Vort}\right) = \vnabla\times \vec{\FF}\,.
\end{equation}

The Helmholtz vorticity conservation expresses the conservation
of angular momentum of fluid particles, and we can equivalently
derive it in its integrated form, namely the conservation of
circulation as first shown by Kelvin.  
We consider a 2-surface $\Sigma$ and define the circulation
$\Circ$ around its boundary $\partial\Sigma$ as
\begin{equation}
  \label{eq:DefCirculation}
  \Circ \equiv \oint_{\partial\Sigma}\form{p} =
  \oint_{\partial\Sigma} p_i \, d x^i\,. 
\end{equation}
Using Stoke's theorem, we see that the circulation around
 $\partial\Sigma$ is equivalent to the vorticity flux through the
 surface $\Sigma$, i.e.
 \begin{equation}
\Circ = \int_{\Sigma} \form{\vort} =  {1\over2}\int_{\Sigma} \vort_{i j}\,
 d x^i\wedge d x^j\,,
\end{equation}
and the more familiar dual expression is found by inserting (\ref{eq:DualVort}):
\begin{equation}
\Circ = \int_{\Sigma} \vec{\Vort}\cdot d\vec{S}\,,  
\end{equation}
where the surface normal element $d\vec{S}$ is 
\mb{$d S_i \equiv {1\over2} \eps_{i j k}\,  d x^j \wedge d x^k$}.
Using (\ref{eq:Helmholtz1}) the comoving time derivative of the
circulation $\Circ$ yields
\begin{eqnarray}
  {d \Circ \over d t}&=& {d\over d t}\oint_{\partial\Sigma}
  \form{p}=  \oint (\partial_t + \Lie_\vv) \, \form{p} 
  = \oint \form{\FF}\,,
  \label{eq:Kelvin}
\end{eqnarray}
which is known as Kelvin's theorem of conservation of circulation.
As we have already seen before, strict conservation only applies if
the non-conservative force per particle $\form{\FF}$ vanishes. 

\subsection{Vorticity and superfluids}
\label{sec:Superfluidity}

The hydrodynamics of superfluids is characterized by two fundamental
properties: on one hand by the absence of dissipative mechanisms like
friction or viscosity, and on the other hand by irrotational flow. As
we will see now, the hydrodynamic description of superfluids is
therefore a natural subclass within the more general framework of
multi-constituent hydrodynamics presented here.  
Let us assume that a constituent $\X= \S$ is superfluid, with 
particle density $n_\S$, velocity $\vv_\S$ and mass $m^\S$.
The absence of microscopic dissipative mechanisms implies that the
superfluid is not bound to any other constituents , i.e. it is a
perfect conductor in the sense that it can flow freely even in the
presence of other constituents. Dissipation-free flow is characterized
by the absence of non-conservative forces acting on the bulk\footnote{
However, there \emph{can} be a non-conservative force acting on the
superfluid at a vortex-core if the vortex is pushed by another
fluid. This mechanism gives rise to the so-called effect of ``mutual
friction''.}  
of superfluid, i.e.  
\begin{equation}
\vec{\FF}^\S =0\,.
\label{eq:SfForceFree}
\end{equation}
As a consequence of (\ref{eq:HelmholtzIa}) and (\ref{eq:Kelvin})   
we see that the vorticity (and therefore circulation) of a superfluid
is strictly conserved.
The second constraint, which distinguishes a superfluid from a perfect
fluid, is that a superfluid is locally \emph{irrotational}, i.e. its
vorticity is zero,  so
\begin{equation}
  \label{eq:IrrotFlow}
  \form{\vort}^\S = 0\,,\equi
  \vec{\Vort}^\S = 0\,.
\end{equation}
Due to the vorticity conservation of superfluids, this constraint
remains automatically satisfied if it is true at some instant $t$,
i.e. it is consistent with the hydrodynamic evolution.

The formulation most commonly  found in the literature on superfluids and
superconductors is based on the concept of the so-called 
``superfluid velocity'', which is constrained to be irrotational
\cite{landau59:_fluid_mech,tilley90:_super}. If
one interpreted this as the actual transport-velocity $\vv_\S$, such a
constraint would generally not be consistent with the equations of
motion, contrary to the natural conservation  of the \emph{momentum
  vorticity} $\form{\vort}^\S$.
This ``orthodox'' formulation of superfluidity, which goes back to
Landau's two-fluid model for $^4$He, is therefore a rather unfortunate
misinterpretation of physical quantities, as the so-called
``superfluid velocity'' is necessarily to be interpreted as the
rescaled \emph{superfluid momentum} in order to make this constraint
consistent with hydrodynamics.
The fact that in Newtonian single-fluid contexts the particle momentum
only differs by a constant mass factor from the velocity has
unfortunately lead to a less than careful distinction between these
fundamentally different quantities.   
This simple identification no longer holds true in more general
contexts, like in the case of multi-fluids (e.g. superfluids) or even
in the case of a single relativistic perfect fluid.
The velocity-circulation is generally \emph{not} conserved, contrary
to the conservation of momentum circulation (\ref{eq:Kelvin}).
The orthodox framework of superfluid hydrodynamics will be discussed
in more detail in Sect.~\ref{sec:He4}.

In addition to the  superfluid constraints of being dissipation-free
and irrotational, there is a further important restriction, namely the
quantization of circulation. 
An irrotational flow can still carry non-zero circulation in the
presence of \emph{topological defects} (such as vortices). In
order to see this, we note that (as a consequence of
(\ref{eq:IrrotFlow})) we can write the superfluid momentum
$\form{p}^\S$ as the gradient of a \emph{phase} $\varphi$, namely 
\begin{equation}
  \form{p}^\S = \hbar\, d\varphi\,,\quad\textrm{i.e.}\quad
  \vp^\S = \hbar \,\vnabla \varphi\,.
\end{equation}
The circulation (\ref{eq:DefCirculation}) can therefore be non-zero if
$\partial\Sigma$ encloses a topological defect in $\varphi$, i.e. a
region where $\varphi$ (and $\vp^\S$) is not defined, as for example
in the case of flow inside a torus. While in the case of a perfect
irrotational fluid the resulting circulation could have any value, the
superfluid phase $\varphi$ is restricted to change only by a multiple 
of $2\pi$ after a complete tour around the defect.
The resulting circulation is therefore quantized as
\begin{equation}
  \Circ = 2 N \pi \hbar \,,\quad \textrm{with}\quad N \in \mathbb{Z}\,,
\end{equation}
which gives rise to the well-known quantized vortex structure of superfluids.

\subsection{Helicity conservation}

Contrary to the conservation laws derived in the previous sections, which have
been known for more than a century, there is a further conserved quantity
namely the so-called helicity, whose existence in hydrodynamics has
only been pointed out comparatively recently by Moffat \cite{moffat69}. 
This quantity is analogous to the magnetic helicity conservation found
in magneto-hydrodynamics \cite{woltjer58}, and it is related to the
topological structure of the  vorticity, i.e. its ``knottedness''
\cite{moffat92:_helicity}. The relativistic analogue of this
conservation has been shown by Carter
\cite{carter79:_act_gal_nuclei,carter89:_covar_theor_conduc,carter92:_momen_vortic_helic},
and generalizations have been discussed by Bekenstein \cite{bekenstein87:_helicity}.   

We define the helicity 3-form $\form{H}$ (with components
$H_{i j k}$) as the exterior product of the momentum 1-form 
$\form{p}$ with the vorticity 2-form $\form{\vort}$, i.e. 
\begin{equation}
  \label{eq:DefHelicity3Form}
  \form{H} \equiv \form{p}\wedge\form{\vort}\,,
\end{equation}
which in components reads as \mb{$H_{i j k} = 3 p_{[i}\vort_{j k]}$}.
A 3-form in a 3-dimensional manifold is dual to a \emph{scalar}, so we
can define the helicity density $h$ as
\begin{equation}
  \label{eq:HelicityScalar}
  H_{i j k} = h \, \eps_{i j k}\,.
\end{equation}
From the duality relation together with the definition
(\ref{eq:DefHelicity3Form}), we see that the
helicity scalar has the following explicit expression  
\begin{equation}
  h\!=\!{1\over 3!}\eps^{i j k} H_{i j k} 
  = p_i \, {1\over 2}\eps^{i j k} \vort_{j k} =
  \form{p}\,\rfloor\vec{\Vort} = \vp\cdot(\vnabla\times\vp)\,. 
  \label{eq:ExplHel}
\end{equation}
Using (\ref{eq:Helmholtz1}) and  (\ref{eq:HelmholtzIa}), the comoving
time-derivative of $\form{H}$ can be expressed as 
\begin{eqnarray}
  (\partial_t + \Lie_\vv)\, \form{H} &=& 
  \left[ (\partial_t + \Lie_\vv)\,\form{p} \right] \wedge \form{\vort}
  +\form{p} \wedge \left[(\partial_t + \Lie_\vv) \,\form{\vort}\right] \nonumber\\[0.2em]
  &=& (d Q + \form{\FF}) \wedge \form{\vort} + \form{p} \wedge d \form{\FF} \nonumber\\[0.2em]
  &=& d (Q \form{\vort}) 
  + \left[  d(\form{p} \wedge \form{\FF}) + 2d\form{\FF}\wedge \form{p}\right] \,.
\end{eqnarray}
We see that, not surprisingly, the vanishing of the non-conservative
force $\form{\FF}$ is a necessary (albeit not sufficient) condition
for the conservation of helicity. We introduce the total helicity
$\Hel$ of a volume $V$ as 
\begin{equation}
  \label{eq:TotalHelicity}
  \Hel \equiv \int_V \form{H} = \int_V h \, d V\,,
\end{equation}
and, assuming \mb{$\form{\FF}=0$}, we find for the comoving time
derivative of $\Hel$: 
\begin{equation}
  \label{eq:HelicityConservation}
  {d \Hel \over d t} = \int_{V} (\partial_t + \Lie_\vv) \, \form{H}
  = \oint_{\partial V} Q \,\form{\vort}
 = \oint_{\partial V} Q \vec{\Vort} \cdot d\vec{S}\,. 
\end{equation}
The helicity $\Hel$ of a volume $V$ is therefore conserved under
transport by the fluid  if, in addition to $\form{\FF}=0$, the
vorticity $\vec{\Vort}$ vanishes on the surface $\partial V$
surrounding this volume.

\section{Hydrodynamics}
\label{sec:Hydrodynamics}

\subsection{The Lagrangian of hydrodynamics}

In the previous sections we have derived the most general form
of the Euler-Lagrange equations (\ref{eq:GeneralEOM}) associated with
the convective variational principle, together with the force
densities (\ref{equfA}) and energy transfer rates (\ref{equgA}). 
We are now interested in a particular class of Lagrangian densities
$\LagrH$, namely those which describe Newtonian hydrodynamics. 
One can postulate the general form of the hydrodynamic Lagrangian
$\LagrH$ in analogy to canonical particle mechanics as 
\be
\LagrH(n_\X, \vn_\X) \equiv \csum m^\X {\vn_\X^2  \over 2 n_\X} - \E\,,
\label{eq:DefLagrH}
\fe
where $\E$ is a thermodynamic potential related to the internal energy
(or ``equation of state'') of the system. We therefore find the
following general form for the conjugate momenta $p^\X_0$ and $\vp^\X$ 
as defined in Eq.~(\ref{equVarL}): 
\begin{equation}
  -p^\X_0 = {1\over2}m^\X \vv_\X^2 + {\partial \E\over \partial n_\X}\,,\quad
  \vp^\X = m^\X \vv_\X - {\partial \E \over \partial \vn_\X}\,.
\label{eq:HydroMomenta}
\end{equation}
We want to identify these conjugate momenta with the actual
physical energy and momentum per fluid particle, which implies that
under a Galilean boost $-\vV$  inducing the transformations
\begin{equation}
  \label{eq:GalileanBoost}
\vv_\X' = \vv_\X + \vV\,,\quad  n_\X' = n_\X\,,\quad
{\partial_t}'  = \partial_t - \vV\cdot\vnabla\,,
\end{equation}
these momenta should transform (e.g. see
\cite{landau59:_fluid_mech,khalatnikov65:_introd}) as 
\be
  -{p^\X_0}'\!=\! -p^\X_0 + \vV\cdot\vp^\X + {1\over2} m^\X \vV^2\,,\qaq
   \vp^\X{}' = \vp^\X + m^\X \vV\,. \label{eq:pX_boost}
\fe
One can verify that in this case the hydrodynamic force densities
$\vfH^\X$ defined in (\ref{eq:DeffH}) are invariant under Galilean
boosts as one should expect.  
The particle creation rates $\Rate_\X$ defined in (\ref{eq:DefRateX}) are
also Galilean invariant, so that the transformation of the total force
densities $\cf^\X$ of (\ref{equfA}) is seen to be    
\begin{equation}
  {\vcf^\X}{}' = \vcf^\X + \vV\, m^\X \Rate_\X\,.
\end{equation}
The equations of motions of an isolated system, i.e. $\csum \vcf^\X=0$,
are therefore Galilean invariant if and only if the total mass is
conserved, i.e. if 
\be
\csum m^\X \Rate_\X = 0\,.
\label{eq:MassCons}
\fe
By using (\ref{eq:pX_boost}) we can show that the energy transfer
rates (\ref{eq:EnFlux}) transform as 
\begin{equation}
  {g^\X}' = g^\X + \vV\cdot \vcf^\X + m^\X \Rate_\X {V^2 \over 2} \,,
\end{equation}
and due to mass conservation (\ref{eq:MassCons}) the total energy
change rate therefore satisfies 
\begin{equation}
  \csum {g^\X}' = \csum g^\X + \vV\cdot\vcf_\ext\,,
\end{equation}
so that the total energy conservation of an isolated system is
Galilean invariant. 

In general the transformation properties (\ref{eq:pX_boost}) are only
consistent with the conjugate momenta (\ref{eq:HydroMomenta}) if $\E$ is
itself Galilean invariant, which is shown in Appendix~\ref{sec:GalileanE}.
This implies that the velocity dependence of $\E$ can only be of the form 
\begin{equation}
  \E(n_\X, \vn_\X) = \E( n_\X, \vDv_{\X\Y})\,,
  \label{eq:InvariantEOS}
\end{equation}
where $\vDv_{\X\Y}$ is the relative velocity between fluid $\X$ and fluid $\Y$, i.e.
\be
\vDv_{\X\Y} \equiv \vv_\X - \vv_\Y = {\vn_\X \over n_\X} - 
{\vn_\Y  \over n_\Y}\,.
\fe
We note that a function $\E$ of the form (\ref{eq:InvariantEOS})
satisfies the identity   
\begin{equation}
\csum n_\X {\partial \E \over \partial \vn_\X} = 0\,,
\end{equation}
which can be used together with (\ref{eq:HydroMomenta}) to show that
the hydrodynamic momentum density (\ref{eq:Tmunu})  satisfies 
\begin{equation}
  \label{eq:MassCurrent}
  \vJ_\H = \csum n_\X \vp^\X = \csum m^\X \vn_\X  = \vec{\rho}\,,
\end{equation}
i.e. the hydrodynamic momentum density $\vJ_\H$ is equal to the total
mass current $\vec{\rho}$ as a consequence of Galilean invariance.   

In addition to the requirement of Galilean invariance we will restrict
our attention to systems of ``perfect'' multi-constituent fluids in
the sense that their energy function $\E$ is isotropic. This means that
we consider only equations of state of the form  
\begin{equation}
  \label{eq:EOS_GeneralForm}
  \E(n_\X, \vDv_{\X\Y}) = \E(n_\X, \Dv_{\X\Y}^2)\,.
\end{equation}
Summarizing we can now write the hydrodynamic Lagrangian density
(\ref{eq:DefLagrH}) for this class of perfect multi-fluid systems as 
\begin{equation}
  \LagrH(n_\X, \vn_\X) = \csum m^\X {\vn_\X^2 \over 2 n_\X}
  - \E(n_\X, \Dv_{\X\Y}^2)\,.
  \label{equLH}
\end{equation}
It is interesting to note that contrary to the relativistic case, 
which is governed by a fully covariant hydrodynamic Lagrangian density
(e.g. see \cite{carter89:_covar_theor_conduc}), the Newtonian
Lagrangian (\ref{equLH}) is {\em not} strictly Galilean invariant 
because of the kinetic energy term. The violation is sufficiently
weak, however, that is does not affect the Galilean invariance of
the resulting equations of motion.

\subsection{Conjugate momenta and entrainment effect}
The total differential of the energy function $\E(n_\X, \Dv_{\X\Y}^2)$
represents the first law of thermodynamics for the given system, namely
\be
d\E =  \csum \mu^\X \,d n_\X + 
{1\over2} \csum_{\X,\Y}\entr^{\X\Y} \, d\Dv^2_{\X\Y}\,, 
\label{equdE}
\fe
which defines the chemical potentials $\mu^\X$ and the symmetric
\emph{entrainment} matrix $\entr^{\X\Y}$ as the thermodynamical
conjugates to $n_\X$ and $\Dv^2_{\X\Y}$. 
The conjugate momenta (\ref{eq:HydroMomenta}) are therefore explicitly 
found  as  
\bea
\vp^\X &=& m^\X \vv_\X - \csum_\Y {2\entr^{\X\Y}\over n_\X}
\vDv_{\X\Y}\,, \label{equpA}\\ 
-p^\X_0 &=& \mu^\X -  m^\X {v_\X^2 \over 2}  +
\vv_\X\cdot\vp^\X\,. \label{equp0A} 
\fea
The expression (\ref{equpA}) for the momenta in terms of the
velocities is interesting, as it shows that in general the momenta
are not aligned with the respective fluid velocities, which is the
so-called entrainment effect\footnote{
Sometimes also referred to as ``drag'' in the superfluid literature.
But as pointed out in \cite{langlois98:_differ_rotat_superfl_ns}, this 
is rather misleading, as entrainment is a purely non-dissipative
effect, whereas ``drag'' in physics usually refers to a resistive drag.}.
The simple single-fluid case, in which the momentum is just  $\vp = m\vv$,
is only recovered if there is no entrainment between the fluids
(i.e. $\entr^{\X\Y}=0$) or if all constituents move together 
(i.e. $\vDv_{\X\Y}=0$). This phenomenon is well-known (albeit not
under the name entrainment) in solid-state 
physics, for example the electron momentum in a crystal lattice is
connected to its velocity by an \emph{effective mass-tensor} (e.g. see
\cite{ziman72:_princ_solid}).  For a more detailed discussion of the
explicit relation between effective masses and entrainment in a
two-fluid model we refer  the reader to
\cite{prix02:_slow_rot_ns_entrain}.  In the context of superfluid
mixtures the importance of the interaction and the entrainment effect
has first been recognized by Andreev\&Bashkin
\cite{andreev75:_three_velocity_hydro}, although  expressed in the
conceptually more confused orthodox framework of superfluidity. 
Substituting (\ref{equLH}) together with (\ref{equp0A}) and
(\ref{equpA}) into (\ref{equDefPress}), we can now relate the
``generalized pressure'' $\Press$ directly to the energy function
$\E$, namely
\be
\E + \Press = \csum n_\X \mu^\X\,,
\label{equTD}
\fe
and with (\ref{equdE}) the total differential of $\Press(\mu^\X,
\Dv^2_{\X\Y})$ is found as 
\be
d \Press = \csum n_\X \,d\mu^\X - {1\over2}
\csum_{\X,\Y}\entr^{\X\Y} \,d\Dv^2_{\X\Y} \label{equdP}\,.
\fe
We can further express the hydrodynamic force density (\ref{eq:DeffH})
more explicitly as 
\begin{equation}
  \vfH^\X\!=\!n_\X \left( \partial_t\!+\!\vv_\X\!\cdot\vnabla \right)
  \,\vp^\X + n_\X \vnabla \mu^\X - \csum_\Y 2\entr^{\X\Y}
  \Dv_{\X\Y}^j \vnabla {v_{\X}}_j \,,
\end{equation}
and for the conserved hydrodynamic energy density (\ref{eq:T0mu}) we find  
\begin{equation}
  \label{eq:ExplEnergyDensity}
  E_\H = \csum_\X m^\X n_\X {v_\X^2\over2} + \E - \csum_{\X,\Y}
  \entr^{\X\Y}\,\Dv_{\X\Y}^2\,.
\end{equation}
This relation can be used to clarify the physical meaning
of the thermodynamic potential $\E$. One might have 
expected to find the total energy density simply as the sum of kinetic
energies plus $\E$. It is to be noted though that $E_\H$, which
represents the Hamiltonian $\HamilH(n_\X, \vp^\X)$ of the system, is
naturally a function of the fluid momenta $\vp^\X$ as opposed to the
velocities.  Similarly it turns out that in order to find the actual
``internal energy'', we have to construct the thermodynamic potential
that depends on the relative momenta instead of $\Dv_{\X\Y}$. We
therefore define the ``entrained'' relative momenta $\vJ^{\X\Y}$ as 
\be
\vJ^{\X\Y} \equiv 2 \entr^{\X\Y} \vDv_{\X\Y}\,,
\fe
representing the momentum exchange between constituents $\X$ and
$\Y$ due to entrainment, namely by using (\ref{equpA}) the momentum
density of the constituent $\X$ can be written as
\be
n_\X \vp^\X = n_\X m^\X \vv_\X - \csum_\Y \vJ^{\X\Y} \,.
\fe
Using this definition of $\vJ^{\X\Y}$, the first law (\ref{equdE}) now
takes the form
\be
d\E =  \csum \mu^\X \,d n_\X + {1\over2} \csum_{\X,\Y} \vJ^{\X\Y}\, d\vDv_{\X\Y}\,, 
\fe
We can therefore introduce the internal energy density $\Et$ as the 
Legendre transformed (with respect to the momenta $\vJ^{\X\Y}$) of the
energy function $\E$, namely  
\be
\Et (n_\X, \vJ^{\X\Y}) \equiv \E - {1\over 2} \csum_{\X,\Y}
\vJ^{\X\Y}\cdot \vDv_{\X\Y}\,,
\fe
with the associated total differential 
\be
d \Et = \csum \mu^\X \,d n_\X - {1\over2} \csum_{\X,\Y}\vDv_{\X\Y}\, d \vJ^{\X\Y}\,.
\fe
We note that $\E$ and $\Et$ only differ in systems where the
entrainment effect is present. Traditionally the quantity $\Et$ is
what one might call the actual ``internal energy'' density, which is a
function of the momenta, while the conjugate thermodynamic potential
$\E$ does not seem to have a well established name in the literature.
We see that in terms of the internal energy $\Et$, the total
energy density (\ref{eq:ExplEnergyDensity}) does indeed have the
expected form of ``kinetic plus internal'' energy, namely 
\be
  E_\H = \csum_\X m^\X n_\X {v_\X^2\over2} + \Et\,.
\fe

\subsection{Entropy and temperature}
\label{sec:Entropy}

As noted earlier, entropy can be included quite naturally in this
framework as a constituent. The corresponding density and current
are $n_\s=s$ and  \mb{$\vn_\s=s \vv_\s$} in terms of the entropy
density $s$ and its transport velocity $\vv_\s$. The entropy is
naturally mass-less, i.e. \mb{$m^\s=0$}. The thermodynamically
conjugate variable to the entropy (its ``chemical potential'') is the
temperature, i.e. $\mu^\s=T$, so (\ref{equdE}) can be written as 
\be
d\E = T\, d s + \csum_{\X\not=\s} \mu^\X\, d n_\X + 
{1\over2} \csum_{\X,\Y}\entr^{\X\Y} \, d\Dv^2_{\X\Y}\,.
\fe
The thermal momenta \mb{$p^\s_0=\T_0$} and \mb{$\vp^\s=\vec{\T}$} of
the entropy constituent are found from (\ref{equpA}) and
(\ref{equp0A}), namely
\bea
\vec{\T} &=& - \csum_\Y {2\entr^{\s\Y}\over s} \vDv_{\s\Y} \,, \label{eq:sMom}\\
-\T_0 &=& T + \vv_\s\cdot \vec{\T}\,.\label{eq:sMom0}
\fea
We see that although the entropy has zero rest mass, it can acquire
a non-zero dynamical momentum $\vec{\T}$ due to entrainment. This can 
also be interpreted as the entropy having a non-zero ``effective mass''.
The hydrodynamic entropy force density $\vfH^\s$ and energy change
rate $g^\s$ defined in (\ref{eq:DeffH}) and (\ref{eq:EnFlux}) yield
\bea
\vfH^\s\!\!&=&\!s\vnabla T 
\!+\!s\left( \partial_t + \vv_\s\!\cdot\!\vnabla \right)\vec{\T}
\!-\!\csum\!2\entr^{\s\Y} \Dv_{\s\Y}^j \vnabla {v_{\s}}_j \,, 
\label{eq:Entropyf}\\
g^\s\!\!&=&\!\vv_\s \cdot\vfH^\s + \left(T+ \vv_\s\cdot \vec{\T}\right)
\,\Rate_\s\,. 
\label{eq:EntropyEnRate}
\fea
We see that the temperature gradient is a driving force of the entropy
constituent, as would be expected.  
We also recognize the term $T\Rate_\s$ in the expression of the energy
transfer rate $g^\s$, which represents the  heat creation ``$T\,d S$''.

\section{Applications}
\label{sec:Applications}

\subsection{Single perfect fluids}
As the first application of the foregoing formalism, we consider a
single perfect fluid consisting of several comoving 
constituents. This multi--constituent fluid is described by the
densities $n_\X$ which move with a single velocity 
$\vv_\X = \vv$, and so the currents are 
$\vn_\X = n_\X\, \vv$. Obviously all the relative velocities vanish in
this case, i.e. \mb{$\vDv_{\X\Y}=0$}, and therefore there is no
entrainment. Here we will explicitly write the entropy with its
density $s$, and we do not include it in the constituent
index set labelled by $\X$, i.e. $\X\not=\s$. The Lagrangian
(\ref{equLH}) for this system is
\begin{equation}
  \LagrH = \csum m^\X n_\X {\vv^2\over2} - \E(s, n_\X)\,,
\end{equation}
and the energy and pressure differentials (\ref{equdE}) and
(\ref{equdP}) simply read as
\be
d \E\!=\!T\,d s + \csum \mu^\X\, d n_\X,\qaq
d P\!=\!s\, d T + \csum n_\X\, d\mu^\X\,,
\label{equTD1fluid}
\fe 
where in the case of a single fluid, the generalized pressure $\Press$
simply reduces to the usual fluid pressure $P$. The fluid momenta
(\ref{equpA}) and (\ref{equp0A}) are 
\be
\vp\,^\X = m^\X\, \vv\,,\qaq
- \po^\X = \mu^\X + m^\X {v^2\over2}\,,
\label{eq:1FMomentum}
\fe
while for the entropy constituent we have with (\ref{eq:sMom}) and
(\ref{eq:sMom0}): 
\begin{equation}
  \vec{\T} = 0\,,\qaq
-\T_0 = T\,.
\end{equation}
The explicit expression for the force densities (\ref{equfA}) and
energy transfer rates (\ref{eq:EnFlux}) are found as     
\bea
\vcf^\X\!\!&=&\!n_\X m^\X \left( \partial_t + \vv\!\cdot\!\vnabla\right) \vv
+ n_\X \vnabla \mu^\X  + m^\X \Rate_\X\,\vv\,,  \label{equfNonBaro} \\ 
g^\X\!\! &=&\!\vv\cdot\vcf^\X + \Rate_\X \mu^\X - m^\X {v^2\over 2} \,\Rate_\X\,,
 \label{equgNonBaro}\\
\vcf^\s\!\! &=&\! s \vnabla T\,, \label{eq:ThermalForce}\\
g^\s\!\!&=&\!\vv\cdot\vcf^\s + T \Rate_\s\,, \label{eq:ThermalRate}
\fea
If we allow for an external force $\vcf_\ext$ and energy exchange rate
$g_\ext$, the equations of motion (\ref{eq:GeneralEOM}) of the system are 
\begin{equation}
\vcf^\s + \csum \vcf^\X = \vcf_\ext \,,\qaq
g^\s + \csum g^\X = g_\ext\,.  
\end{equation}
Inserting (\ref{equfNonBaro})--(\ref{eq:ThermalRate}) and using mass
conservation (\ref{eq:MassCons}), we find the explicit equations of
motion 
\begin{eqnarray}
\left(\partial_t + \vv\cdot\vnabla\right)\vv + {1\over \rho}\vnabla P
 &=& {1\over \rho} \vcf_\ext\,,\label{eq:EulerEquation}\\
  T \Rate_\s + \csum \mu^\X \Rate_\X &=& g_\ext - \vv\cdot\vcf_\ext
  \,, \label{eq:1FluidEOM0} 
\end{eqnarray}
where we have used the thermodynamic relation (\ref{equTD1fluid}) in
order to rewrite the momentum equation in the familiar Euler form.
The energy equation expresses the heat creation $T\Rate_\s$ by
chemical reactions $\Rate_\X$. For an \emph{isolated} system, where
$\vcf_\ext=0$ and $g_\ext=0$,  that entropy can only increase due to
the second law of thermodynamics, so \mb{$\Rate_\s \ge 0$}.
From (\ref{eq:1FluidEOM0}) we therefore obtain a constraint on the
direction of the chemical reactions, namely  
\begin{equation}
  \csum \Rate_\X \mu^\X \le 0\,.
\end{equation}
If we consider for example the case of two constituents of equal
mass, so that the mass-conservation (\ref{eq:MassCons}) implies
$\Rate_1 + \Rate_2 = 0$, then this constraint now reads as 
\begin{equation}
  \Rate_1 (\mu^1 - \mu^2) \le 0\,,
\end{equation}
which shows that chemical reactions only proceeds in the direction of
the lower chemical potential as would be expected.  

\subsection{``Potential vorticity'' conservation: Ertel's theorem}

We now consider the case without chemical reactions, in which the
general perfect fluid discussed in the foregoing section can be
described effectively as a fluid consisting only of a single matter
constituent and entropy. 
In this case we can show that the vorticity is generally not conserved,
but that a weaker form of the vorticity conservation still holds.
The fluid is described by the particle number density $n$, the mass
per particle $m$ and a comoving entropy density $s$. 
Mass conservation (\ref{eq:MassCons}) in this case reduces to $\Rate=0$.
If we assume the system to be isolated, i.e. \mb{$\vec{f}+\vec{f}^\s=0$},
then the only force per particle (\ref{eq:ForceSeparation}) acting on the
matter constituent is the ``thermal force'' (\ref{eq:ThermalForce}), 
namely 
\begin{equation}
  {1\over n}\vfH = - \st \,\vnabla T\,,
\label{eq:Thermalpfpp}
\end{equation}
where $\st \equiv {s / n}$ is the specific entropy. If $\st$ is constant
everywhere, then this ``thermal force'' is conservative, i.e. $\vec{\FF}=0$
and by (\ref{eq:Kelvin}) the circulation is therefore conserved. In
the non-uniform case, however, we find 
\begin{equation}
  {d \Circ \over d t} = \oint_{\partial\Sigma} \form{\FF} = 
  -\oint_{\partial\Sigma} \st \,d T\,, 
\end{equation}
which vanishes only if we integrate along a path $\partial\Sigma$ that
lies completely in a surface of constant $\st$. 
We can also see this in the Helmholtz formulation, namely by applying
an exterior derivative to (\ref{eq:Thermalpfpp}), one obtains
\begin{equation}
  d\form{\FF} = - d\st \wedge d T\,,\quad\textrm{i.e.}\quad
  \vnabla\times\vec{\FF} = -\vnabla\st \times \vnabla T\,,
\end{equation}
and it follows therefore from (\ref{eq:HelmholtzIa}) that the
vorticity is no longer generally conserved in this case. However, 
the quantity $d\st\wedge d\form{\FF}$, or its equivalent dual expression
\mb{$\vnabla\st\cdot(\vnabla\times\vec{\FF})$}, still vanishes
identically. 
Based on this observation we construct the ``potential vorticity''
3-form $\form{\Z}$ as 
\begin{equation}
  \label{eq:DefPotVortForm}
  \form{\Z} \equiv d\st \wedge \form{\vort}\,,
\end{equation}
and the dual scalar $\z$ is
\begin{equation}
  \Z_{i j k} = \z \,\eps_{i j k}\,,\qaq 
\z = {1\over 3!} \eps^{i j k} \Z_{i j k} = \vnabla \st\cdot
(\vnabla\times\vp) \,,  
\label{eq:DefPotVortScalar}
\end{equation}
where the last expression was found using (\ref{eq:DualVort}).
The evolution of the potential vorticity 3-form $\form{\Z}$ under transport by
the fluid is
\begin{equation}
  (\partial_t + \Lie_v) \,\form{\Z} = 
d\left[ (\partial_t + \Lie_v) \st\right] \wedge \form{\vort}\,,
\label{eq:Ertel1}
\end{equation}
and therefore $\form{\Z}$ is conserved for isentropic flow, i.e. if
\begin{equation}
\Rate_\s = 0 \equi  (\partial_t + \Lie_v)\,\st = 0\,. 
\end{equation}
The dual version of (\ref{eq:Ertel1}), namely the conservation of the
scalar $\z$ is then found as
\begin{equation}
  \label{eq:Ertel2}
  \partial_t \,\z + \vnabla\cdot(\z\vv) = 0\,.
\end{equation}
Traditionally this conservation law is often expressed in terms of the
scalar $\alpha \equiv z/\rho$, which then results in the following
form of the conservation law:
\begin{equation}
  \label{eq:ErtelTrad}
  \left(\partial_t + \vv\cdot\vnabla \right) \, \alpha = 0\,,
\end{equation}
which is generally known as ``Ertel's theorem''
\cite{ertel42,schutz80:_geomet}. 

\subsection{Thermally conducting fluids}

We have so far only considered perfect fluids, which are perfect
heat insulators as the entropy is strictly carried along by fluid
elements and no heat is exchanged between fluid elements.
It is quite straightforward to extend this to thermally conducting fluids
simply by dropping the assumption that the entropy flux is bound to
the matter fluid flow, i.e. we just have to allow  
\mb{$\vv_\s \not=\vv$}, where $\vv_\s$ and $\vv$ are the velocities of
the entropy fluid and the matter fluid respectively. For simplicity we
consider only a single matter constituent, described by its particle
number density $n$, which by (\ref{eq:MassCons}) is automatically
conserved, i.e. $\Rate=0$.  

From the general expressions (\ref{eq:sMom0}) and (\ref{eq:sMom}) we
see that the ``entropy fluid'' acquires a non-zero momentum
due to the interaction with the matter fluid, via entrainment. 
However, this aspect does not usually seem to be taken into account in
the traditional description of heat-conducting fluids (e.g. see
\cite{landau59:_fluid_mech}). The aim of the present section is only
to show how to recover the standard equations for a
heat-conducting fluid, and we therefore simply assume the entrainment
to be negligible, i.e. $\entr=0$.  It is certainly an interesting
question if this neglect of entrainment is physically justified in all
cases. With this assumption, the force density (\ref{eq:Entropyf}) and
energy rate (\ref{eq:EntropyEnRate}) of the entropy reduce to
\begin{equation}
  \vcf^\s = s \vnabla T \,,\qaq
  g^\s = \vv_\s\cdot\vcf^\s + T \, \Rate_\s\,.
  \label{eq:EntropyForce}
\end{equation}
As in the (isolated) perfect fluid case discussed previously, the
equations of motion are again \mb{$\vcf^\s + \vcf = 0$} and \mb{$g^\s
  + g = 0$}.  
This time, however, one force density, $\vcf^\s$ say, can be specified
by the model due to the increased number of degrees of freedom, so we
set it to \mb{$\vcf^\s=\vcf_\R$}, where $\vcf_\R$ is a resistive force
acting against the entropy flow.
We obtain the Euler equation in the same form as
in (\ref{eq:EulerEquation}), but now the energy equation takes the
form
\begin{equation}
  T \Rate_\s = (\vv - \vv_\s) \cdot \vcf_\R\,.
\end{equation}
By the second law of thermodynamics, namely $\Rate_\s\ge0$, we can
constrain the form of the resistive force $\vcf_\R$ to
\begin{equation}
  \vcf_\R = -\eta\,( \vv_\s - \vv)\,,\quad\textrm{with}\quad \eta \ge 0\,,
  \label{eq:EntropyFriction}
\end{equation}
i.e. the friction force acting on the entropy fluid is always opposed
to its flow relative to the matter fluid. Obviously the
value of the resistivity $\eta$ is not restricted to be a constant
but will generally depend on the state of the system. 
Following the traditional description
(e.g. \cite{landau59:_fluid_mech}) we introduce the heat flux
density $\vq$ \emph{relative to the matter fluid} as
\begin{equation}
  \label{eq:DefHeatFlux}
  \vq \equiv T s ( \vv_\s - \vv) \,.
\end{equation}
By combining this with (\ref{eq:EntropyForce}) and
(\ref{eq:EntropyFriction}), we see that the heat flux current
is constrained by the second law to be of the form
\begin{equation}
  \vq = -\kappa \vnabla T\,,\quad\textrm{with}\quad
  \kappa \equiv {T s^2 \over \eta} \ge 0\,,
\end{equation}
where $\kappa$ is the \emph{thermal conductivity}.
With (\ref{eq:DefHeatFlux}) we can express the velocity of the entropy
fluid $\vv_\s$ in terms of the heat flux $\vq$ and the matter
velocity $\vv$, so the entropy creation rate $\Rate_\s$ can be
expressed as
\begin{equation}
  \Rate_\s = \partial_t s + \vnabla\cdot\left( s \vv + {\vq\over T} \right)\,.
\end{equation}
We further find for the hydrodynamic energy flux vector $\vQ_\H$ of
(\ref{eq:T0mu}):
\begin{eqnarray}
  \vQ_\H &=& \csum (-p^\X_0) \vn_\X  = (\mu + m {v^2\over2}) n \vv +
  s T \vv_\s\nonumber\\
  &=& n \vv \left(m{v^2\over2} +  \mu + \st T \right) + \vq\,,
\end{eqnarray}
where the last equality was found using (\ref{eq:DefHeatFlux}).
We introduce the specific enthalpy as \mb{$\enth \equiv \mu + \st\,T$}, and
using the first law\footnote{In the absence of entrainment the
  entropy fluid does not carry momentum, therefore the matter fluid
  defines a unique frame in which the stress tensor (\ref{eq:Tmunu})
  is purely isotropic. In this case the generalized pressure $\Press$
  is identical with the usual perfect fluid notion of the pressure
  $P$.}, namely \mb{$d P = n\,d\mu + s\,d T$}, we find the total
variation of the specific enthalpy as 
\begin{equation}
  d \enth = T d \st + {1\over n} d P\,,
\end{equation}
and so we recover the standard expression
(e.g. cf. \cite{landau59:_fluid_mech}) for the energy flux:
\begin{equation}
  \vQ_\H = n \vv \left( m {v^2\over2} + \enth \right) + \vq\,.
\end{equation}

\subsection{The two-fluid model for superfluid $^4$He}
\label{sec:He4}

We now consider the example of superfluid $^4$He at a non-zero
temperature $T$. Let $n$ be the number density of $^4$He atoms and $s$
be the entropy density. The $^4$He atoms move with a velocity $\vv$, while
the entropy (carried by a thermal gas of excitations such as phonons
and rotons) transports heat without friction (i.e. $\vcf_\R=0$) at the
velocity $\vv_\N$, so the relative velocity is 
\mb{$\vDv = \vv_\N - \vv$}. In this context the entropy fluid is
often referred to as the ``normal fluid'' as opposed to the superfluid
mass flow. The two transport currents, namely that of $^4$He atoms and
of entropy,  are respectively 
\begin{equation}
  \vn = n\, \vv\,,\qaq
  \vec{s} = s \, \vv_\N\,.
  \label{eq:He4ConservedCurrents}
\end{equation}
The $^4$He atoms have mass $m$, so the mass density is $\rho = n m$, and
the hydrodynamic Lagrangian density (\ref{equLH}) reads as
\begin{equation}
  \LagrH = {1\over 2} n m v^2 - \E(n, s, \Dv^2)\,,
\end{equation}
where the energy function $\E$ determines the first law (\ref{equdE}) as
\begin{equation}
  d\E = \mu\,d n + T \,d s + \entr \, d \Dv^2\,,
  \label{eq:HeEnergyVar}
\end{equation}
which defines the chemical potential $\mu$ of $^4$He atoms, the
temperature $T$ and the entrainment $\entr$. The conjugate momenta
(\ref{equpA}), (\ref{equp0A}) of the $^4$He atoms are
\begin{eqnarray}
  \vp &=& m \vv + {2\entr \over n}\vDv\,, \label{eq:He4vp}\\
  -\po &=& \mu -{1\over 2}m v^2 +  \vv\cdot\vp\,,
  \label{eq:He4p0}
\end{eqnarray}
while for the entropy fluid Eqs.~(\ref{eq:sMom}) and (\ref{eq:sMom0})
yield 
\begin{eqnarray}
  \vec{\T} &=& -{2\entr\over s} \vDv\,,\label{eq:Entropyvp}\\
  -\T_0 &=& T + \vv_\N\cdot \vec{\T}\,. \label{eq:Entropyp0}
\end{eqnarray}
The conservation of mass (\ref{eq:MassCons}) implies 
\begin{equation}
  \Rate = \partial_t n + \vnabla\cdot\vn = 0\,.
\end{equation}
In the absence of vortices, there are no direct forces acting between
the two fluids, so the equations of motion in the absence of external
forces (i.e. \mb{$\vcf_\ext=0$}) are simply
\begin{equation}
  \vf = \vfH = 0\,\qaq \vcf^\N = 0\,.
\label{eq:He4EOM0}
\end{equation}
The energy equations are $g=0$ and $g^\N = g_\ext$, and with
(\ref{eq:EntropyEnRate}) this leads to
\begin{equation}
 -g_\ext =  \Rate_\s ( \T_0 + \vv_\N\cdot \vec{\T}) = -T \Rate_\s \,,
\end{equation}
where we have inserted (\ref{eq:Entropyp0}). We see that this equation
describes the rate of entropy creation by an external heat source, namely
\begin{equation}
  \partial_t s + \vnabla\cdot(s \vv_\N) = {1\over T}g_\ext\,.
\end{equation}
As discussed in Sect.~\ref{sec:Superfluidity}, the superfluid $^4$He
is (locally) irrotational, i.e.
\begin{equation}
  \vort_{i j} = 2 \nabla_{[i} p_{j]} = 0\,,\Longleftrightarrow
\vec{\Vort} = \vnabla\times \vp = 0\,.
\end{equation}
Using (\ref{eq:DeffH}), the equation of motion (\ref{eq:He4EOM0}) for
the superfluid therefore  reduces to  
\begin{equation}
  \partial_t \vp - \vnabla \po = 0\,,
  \label{eq:SfAccel}
\end{equation}
and with the explicit momenta (\ref{eq:He4p0}) and (\ref{eq:He4vp})
this yields
\begin{equation}
  \partial_t \left( \vv + \entrn \vDv \right) + \vnabla \left(  
    \mut + {1\over2} v^2 + \entrn \vv\cdot \vDv \right)   = 0\,, 
\label{eq:He4EOM1}
\end{equation}
where we introduced the entrainment number $\entrn$ and the specific
chemical potential $\mut$ as
\begin{equation}
  \entrn \equiv {2\entr \over \rho}\,,\qaq
  \mut \equiv {\mu \over m }\,.
  \label{eq:DefEpsMut}
\end{equation}
The  entropy fluid is governed by the
momentum equation \mb{$\vcf^\N=0$}, and with (\ref{eq:Entropyf}) and
the entropy momenta (\ref{eq:Entropyp0}) and (\ref{eq:Entropyvp}), we find
\begin{equation}
  (\partial_t + \vv_\N\cdot\vnabla) \left( {2\entr\over s}\vDv
  \right) - \vnabla T + {2\entr \over s} \Dv_j \vnabla v_\N^j 
+ {2\entr \over s^2} \Rate_\s \vDv = 0\,.
  \label{eq:He4EOM2}
\end{equation}
The two equations (\ref{eq:He4EOM1}) and (\ref{eq:He4EOM2})
represent the ``canonical'' formulation of the two-fluid model for
superfluid $^4$He. These equations do not seem to bear any obvious
relation to the ``orthodox'' formulation of Landau's two-fluid model
found in all textbooks on the subject (e.g. see  
\cite{landau59:_fluid_mech,khalatnikov65:_introd,tilley90:_super}).
Nevertheless, these equations are equivalent to the orthodox
framework, as we will show now, but it is important to note that the
orthodox formulation is based on a rather unfortunate confusion
between the velocity and momentum of the superfluid which is inherent
in the historic definition of the ``superfluid velocity'' by Landau. 

We demonstrate the equivalence of these formulations by explicitly
translating the canonical formulation into the orthodox language. 
The starting point of Landau's model is the statement that the
``superfluid velocity'' is irrotational. We write $\vec{\V}_\S$ for
the ``superfluid velocity'', which is not to be confused with the
actual velocity $\vv$ of $^4$He atoms, so the starting point is  
\begin{equation}
  \vnabla \times \vec{\V}_\S = 0\,.\label{eq:LandauIrrot}
\end{equation}
From the general discussion about vorticity conservation in
Sect.~\ref{sec:Vorticity} and its particular role in superfluids
(Sect.~\ref{sec:Superfluidity}) we have already seen that contrary to
the momentum vorticity \mb{$\vec{\Vort}=\vnabla\times\vp$}, the
velocity-rotation \mb{$\vnabla\times\vv$} is generally \emph{not}
conserved by the fluid flow, and in particular not in the presence
of more than one fluid as is the case in superfluid $^4$He at $T>0$. 
The only possible interpretation we can give $\vec{\V}_\S$ in order
for the constraint (\ref{eq:LandauIrrot}) to be consistent with
hydrodynamics and to remain true for all times is that it is really
the rescaled superfluid \emph{momentum} $\vp$, so the ``key'' to our
translation is the ansatz 
\begin{equation}
  \vec{\V}_\S \equiv {\vp \over m}\,.
\end{equation}
While this would be equivalent to the fluid velocity in a single
perfect fluid, as seen in (\ref{eq:1FMomentum}), this has no
interpretation as the velocity of either the mass or the entropy
in the case of the present two-fluid model as we can see in
(\ref{eq:He4vp}). Therefore we call $\vec{\V}_\S$ a \emph{pseudo velocity},  
as it is a \emph{dynamic} combination of both fluid velocities
governed by the entrainment $\entr$ between the superfluid $^4$He and
its excitations.  
With the explicit entrainment relation (\ref{eq:He4vp}) we can now
express the velocity $\vv$ of the $^4$He fluid in terms of the
pseudo-velocity $\vec{\V}_\S$ and the normal-fluid velocity $\vv_\N$ as
\begin{equation}
  \vv = (1-\entrn)^{-1}\, ( \vec{\V}_\S - \entrn \vv_\N)\,,
  \label{eq:RelVsVel}
\end{equation}
where we used the definition (\ref{eq:DefEpsMut}) of the entrainment
number $\entrn$.
With this substitution, the total mass current $\vec{\rho}$, which is
equal to the total momentum density $\vJ_\H$ as seen in
(\ref{eq:MassCurrent}), can be written in the form
\begin{equation}
  \vJ_\H = \rho \vv = \left[{\rho \over 1-\entrn}\right] \,\vec{\V}_\S
  + \left[{-\entrn \rho \over 1 - \entrn}\right] \, \vv_\N\,,
\end{equation}
which suggests to introduce a ``superfluid density'' $\prho_\S$ and
a ``normal density'' $\prho_\N$ as 
\begin{equation}
  \prho_\S \equiv {\rho \over 1 - \entrn}\,,\qaq 
  \prho_\N \equiv {-\entrn \rho \over 1 - \entrn}\,,
  \label{eq:DefPseudoDens}
\end{equation}
such that total mass density $\rho$ and mass current
$\vec{\rho}=\vJ_\H$ can now be written as
\begin{equation}
  \rho = \prho_\S + \prho_\N\,,\qaq
  \vJ_\H = \prho_\S \vec{\V}_\S + \prho_\N \vv_\N\,.
  \label{eq:PseudoDensCur}
\end{equation}
It is now obvious that this split is completely artificial,
and $\prho_\N$ and $\prho_\S$ are only \emph{pseudo densities},
as they do not represent the density of any (conserved) physical
quantity and are not even necessarily positive. In fact neither of the
two pseudo-densities and currents are conserved individually, contrary
to the physical currents (\ref{eq:He4ConservedCurrents}). 
We note that even Landau warned against taking too literally the
interpretation of superfluid $^4$He as a ``mixture'' of these two
(pseudo-) ``fluids'' \cite{landau59:_fluid_mech}. Contrary to the
artificial orthodox split, however, the separation into entropy fluid
and $^4$He mass flow is physically perfectly meaningful, and the
superfluid \emph{can} be regarded as a two-fluid system in the literal
sense in the canonical framework.
The pseudo ``mass density'' $\prho_\N$, which the normal fluid
seems to carry in the orthodox description is due to the fact that
entrainment provides the entropy fluid with a non-vanishing
\emph{momentum} (\ref{eq:Entropyvp}) in the presence of relative
motion, even though it does not transport any mass.  
This lack of careful distinction between mass current and momentum
leads to the paradoxical picture of the ``superfluid counterflow'': 
for example, in the simple case of heat flow through a static 
superfluid, the normal fluid associated with the heat flow carries
a pseudo mass-current $\prho_\N \vv_\N$. But because there is no net 
mass current there has to be some superfluid ``counterflow'' of
pseudo mass current $\prho_\S \vec{\V}_\S = - \prho_\N \vv_\N$.
This apparently strange behavior is solely due to an awkward choice
of variables and a loss of direct contact between the quantities used
in the orthodox description and the actual conserved physical
quantities of $^4$He.  

Further following the traditional orthodox framework, we define
the relative (pseudo-)velocity $\pDv$ as 
\begin{equation}
  \pDv \equiv \vv_\N - \vec{\V}_\S\,,
\end{equation}
which, using (\ref{eq:RelVsVel}), can be expressed in terms of $\vDv$ as
\begin{equation}
  \pDv = (1-\entrn)\, \vDv\,.
\end{equation}
In order to relate the canonical thermodynamic quantities to the
orthodox language, we follow Khalatnikov
\cite{khalatnikov65:_introd} and Landau \cite{landau59:_fluid_mech} 
and consider the energy density in the ``superfluid frame'' $K_0$,
which is defined  by \mb{$\vec{\V}_\S^\sff=0$}.
In this frame, the momentum density $\vJ_\H^\sff$ expressed in
(\ref{eq:PseudoDensCur}) is
\begin{equation}
  \vJ_\H^\sff = \prho_\N \, \vv_\N^\sff = \prho_\N \, \pDv = 
- 2\entr\, \vDv\,,
\end{equation}
and the transport velocity $\vv$ of the superfluid $^4$He atoms in this
frame can be expressed using (\ref{eq:ExprVel}) as
\begin{equation}
  \vv^\sff = \vv - \vec{\V}_\S = {\prho_\N \over \rho} \vv_\N^\sff = 
{1\over \rho} \vJ_\H^\sff\,. 
\label{eq:ExprVel}
\end{equation}
The hydrodynamic energy density $E_\H$ of the fluid system  is given by
(\ref{eq:ExplEnergyDensity}), which reads in this case
\begin{equation}
  E_\H = {1\over 2}\rho \vv^2 + \E - 2 \entr\, \vDv^2\,, 
\end{equation}
and using the previous translations together with the first law
(\ref{eq:HeEnergyVar}), we can write the total variation $d E^\sff$ of
the energy density in $K_0$ as 
\begin{equation}
  \label{eq:OrthEnergyVar}
  d E_H^\sff = T\, d s + \mut_\S \, d\rho + \pDv\cdot d \vJ_\H^\sff\,,
\end{equation}
which defines the ``superfluid chemical potential'' $\mut_\S$ as
\begin{equation}
  \mut_\S = \mut - {1\over2} (\vv - \vec{\V}_\S)^2\,.
\end{equation}
Using these quantities, the canonical equation of motion
(\ref{eq:He4EOM1}) can now be translated into the orthodox
form as
\begin{equation}
  \label{eq:He4OrthEOM1}
  \partial_t \vec{\V}_\S + \vnabla \left( {\V_\S^2\over 2} +
  \mut_\S\right) = 0\,.
\end{equation}
One can equally verify that the generalized pressure, defined in
(\ref{equTD}), is expressible in terms of the orthodox quantities as
\begin{equation}
  \Press = - \E + \rho\, \mut + s\, T = - E_\H^\sff + T\,s +
  \rho\,\mut_\S + \pDv\cdot \vJ_\H^\sff\,,
\end{equation}
in exact agreement with the expressions found in
\cite{khalatnikov65:_introd,landau59:_fluid_mech}.
For the remaining momentum equation, the total momentum conservation
(\ref{eq:MomCons}) is traditionally preferred over the equation
of motion (\ref{eq:He4EOM2}) of the entropy fluid.
We therefore conclude this section by the appropriate translation
of the stress tensor (\ref{eq:Tmunu}) into the orthodox language. 
The canonical expression for the stress tensor of superfluid $^4$He is
\begin{equation}
  T_\H^{i j} = n^i \, p^j + s^i \, \T^j + \Press\, g^{i j}\,,
\end{equation}
and inserting the previous expressions for the explicit momenta and
the translations to orthodox variables, one can write this in
the form 
\begin{equation}
  T_\H^{ij} = \prho_\S \, \V_\S^i \V_\S^j + \prho_\N\, v_\N^i v_\N^j +
  \Press\, g^{i j}\,,
\end{equation}
which concludes our proof of equivalence between canonical and
orthodox description.

\subsection{A two-fluid model for the neutron star core}

Here we consider a (simplified) model for the matter inside a neutron
star core, which mainly consists of a (charge neutral) plasma of
neutrons, protons and electrons. 
We focus on superfluid models, in which the neutrons are
assumed to be superfluid, which allows them to freely traverse
the fluid of charged components due to the absence of viscosity. As
discussed in Sect.~\ref{sec:Superfluidity}, this also implies some
extra complications due to the quantization of vorticity into
microscopic vortices. Here we are interested in a macroscopic
description, i.e. we consider fluid elements that are small compared
to the dimensions of the total system, but which contain a large
number of vortices. On this scale we can work with a smooth averaged
vorticity instead of having to worry about individual vortices. One
effect of the presence of the vortices will be a slight anisotropy in
the resulting smooth averaged fluid  
\cite{bekarevich61:_phenom_vortex,carter95:_kalb_ramond,carter98:_relat_supercond_superfl},
which can be ascribed to the tension of vortices, and which we will
neglect here for simplicity. The second effect of the vortex lattice
is that it allows a direct force between the superfluid and the
normal fluid, mediated by the respective vortex interactions, and
which is naturally described in the context of the two-fluid model as
a mutual force. The model assumptions used here are fairly common
to most current studies of superfluid neutrons stars (e.g. see 
\cite{mendell91:_coupl_charg,lindblom94:_oscil_superfl_ns,andersson01:_dyn_superfl_ns,prix02:_slow_rot_ns_entrain}).

The model therefore consists of comoving constituents 
\mb{$\X\in \{\e,\p, \s\}$}, corresponding to the electrons, protons
and entropy, and   we will label this fluid with '$\c$'. The
second fluid consists only of the superfluid neutrons, i.e. 
\mb{$\X = \n$}. 
Charge conservation implies 
\be
\Rate_\e = \Rate_\p\,,
\label{eq:Rate1}
\fe
and for simplicity we will assume local {\em charge neutrality}, i.e. 
\be
n_\e = n_\p\,.
\fe
We assume the electrons and protons to be strictly moving together in
this model (i.e. we consider timescales longer than the plasma
oscillation timescale), so we can neglect electromagnetic interactions
altogether.  Another physical constraint is {\em baryon conservation},
i.e. we must have 
\be
\Rate_\n + \Rate_\p = 0\,,
\label{eq:Rate2}
\fe
and together with mass conservation (\ref{eq:MassCons}), this leads to
the requirement\footnote{ This relation is of course not exactly 
satisfied in reality, which shows a well-known shortcoming of
Newtonian physics: mass has to be conserved separately from energy.}
\be
m^\n = m^\p + m^\e \equiv m\,.
\fe
We can therefore write the mass densities of the two fluids as 
\begin{equation}
  \rho_\n = m \,n_\n\,,\qaq \rho_\c = m \,n_\p\,.
\end{equation}
The first law (\ref{equdE}) of this model reads as
\bea
d \E\!\!&=&\! T\, d s + \mu^\n\, d n_\n + \mu^\e\, d n_\e + \mu^\p \, d n_\p + 
\entr^{\e\n}\, d\Dv^2_{\e\n}\nonumber\\
&&  + \entr^{\p\n}\, d\Dv^2_{\p\n} + \entr^{\s\n}\,d\Dv^2_{\s\n}\,.
\fea
Obviously there is only one independent relative velocity $\vDv$, namely
\begin{equation}
\vDv \equiv \vv_\c - \vv_\n = \vDv_{\e\n} = \vDv_{\p\n} = \vDv_{\s\n}\,,
\end{equation}
and we define the total entrainment $\entr$ as
\be
\entr \equiv \entr^{\e\n} + \entr^{\p\n} + \entr^{\s\n}\,.
\fe
In the case of the neutron star model, we are obviously also
interested to include the effects of gravitation. We can therefore not
assume the system to be isolated and we include the effect of the
gravitational potential $\Phi$ as an external force. The minimal
equations of motion (\ref{eq:GeneralEOM}) therefore read as
\begin{equation}
  \vcf^\n + \vcf^\c = - \rho \vnabla \Phi\,,\qaq
  g^\n + g^\c = - \vec{\rho}\cdot\vnabla\Phi\,,
  \label{eq:NSEOM0}
\end{equation}
where the force and energy rate of the '$\c$'-fluid are naturally
given by \mb{$\vf^\c \equiv \vf^\p + \vf^\e + \vf^\s$} and 
\mb{$g^\c \equiv g^\p + g^\e + g^\s$}.
With (\ref{eq:Rate1}) and (\ref{eq:Rate2}) we can write the respective
force densities more explicitly as  
\begin{eqnarray}
  \vf^\n &=& \vfH^\n + \Rate_\n \vp^\n\,, \label{eq:NSForce1}\\
  \vf^\c &=& \vfH^\c  - \Rate_\n ( \vp^\e +
  \vp^\p) + \Rate_\s \vec{\T}\,, \label{eq:NSForce2} 
\end{eqnarray}
where we naturally defined \mb{$\vfH^\c\equiv \vfH^\p + \vfH^\e + \vfH^\s$}.
Similarly we can write the energy rates (\ref{eq:EnFlux}) as
\begin{eqnarray}
  g^\n  &=& \vv_\n \cdot \vfH^\n - \Rate_\n \po^\n\,,\label{eq:NSEnRate1}\\
  g^\c  &=& \vv_\c \cdot \vfH^\c + \Rate_\n (\po^\e + \po^\p) - \Rate_\s\T_0 \,. \label{eq:NSEnRate2}
\end{eqnarray}
Because the gravitational acceleration is the same for all bodies
(i.e. fluids), we can now simply absorb the effect of the
gravitational potential into the definition of ``extended'' forces
$\vef$ and energy rates $\eg$ which simply incorporate the respective
gravitational force density and work rate, i.e. we define
\bea
\vef^\X  &\equiv& \vf^\X + \rho_\X \vnabla \Phi\,,\\
\vefH^\X &\equiv& \vfH^\X + \rho_\X \vnabla \Phi\,,\\
\eg^\X   &\equiv& g^\X + \rho_\X \vv_\X\cdot\vnabla\Phi\,.
\fea
With these redefinitions, the minimal equations of motion
(\ref{eq:NSEOM0}) again take the form of an isolated system, i.e.
\be
  \vef^\n + \vef^\c = 0\,,\qaq \eg^\n + \eg^\c = 0\,,
  \label{eq:NSEOM}
\fe
while for (\ref{eq:NSForce1})--(\ref{eq:NSEnRate2}) we obtain exactly
the same form, just for all forces and energy rates replaced by their
``extended'' version. Using the foregoing equations, we obtain
\be
\vefH^\c = -\vef^\n + \Rate_\n \vp^\c - \Rate_\s \vec{\T}\,, \label{eq:NSefHc}
\fe
and therefore
\be
\eg^\c = -\vv_\c\cdot \vefH^\n 
- \Rate_\n \left[\vv_\c\cdot(\vp^\n - \vp^\c) - \po^\c \right] -
\Rate_\s \T_0\,.
\fe
Substituting this and the ``extended'' version of (\ref{eq:NSEnRate1})
into the energy-rate equation (\ref{eq:NSEOM}), we find
\be
T\Rate_\s = \vDv\cdot\vefH^\n + \Rate_\n \left[ \po^\n - \po^\e -
  \po^\p + \vv_\c\cdot\left(\vp\,^\n - \vp\,^\e - \vp\,^\p  \right) 
\right]\,,
\label{equNSEnergy}
\fe
where we have used the explicit form (\ref{eq:sMom0}) of $\T_0$.
In addition to the external force, the two-fluid model allows one to
prescribe one of the fluid force densities. In the present case it is
most convenient to specify the ``extended'' hydrodynamic force
$\vefH^\n$ on the neutrons. As this force can only originate from
the second fluid, we will refer to it as the \emph{mutual force}
$\vf_\mutual$, so we set
\begin{equation}
\vefH^\n = \vf_\mutual\,.
\label{eq:NSfmutual}
\end{equation}
Substituting the explicit conjugate momenta (\ref{equpA}) and
(\ref{equp0A}), we obtain the final expression for the entropy
creation rate (\ref{equNSEnergy}) as
\be
T\Rate_\s = \vDv\cdot\vcf_\mutual + \Rate_\n \beta\,. 
\label{eq:EntropyCreation}
\fe
The first term on the right hand side is the work done by the
mutual force, and the second term is the entropy created by
beta reactions between the two fluids, for 
which the term ``transfusion'' has been coined
\cite{langlois98:_differ_rotat_superfl_ns}.  
The deviation from beta equilibrium characterized by $\beta$ is found
as 
\be
\beta \equiv \mu^\p +\mu^\e - \mu^\n - {1\over2}m
\left(1-{4\entr\over\rho_\n}\right) \Dv^2\,, 
\fe
where the last term gives the correction to the chemical equilibrium
due to relative motion $\vDv$ of the two fluids.
The second law of thermodynamics for an isolated system states that
entropy can only increase, i.e. \mb{$\Rate_\s\ge0$}. 
In order for this to be identically true in
(\ref{eq:EntropyCreation}), the mutual force $\vcf_\mutual$ and
the reaction rate $\Rate_\n$ have to be of the form
\be
\begin{array}{r l l l}
\Rate_\n =& \Xi \, \beta \,,&\textrm{with}\quad \Xi \ge 0\,,\\
\vcf_\mutual =& \eta\, \vDv + \vec{\kappa}\times \vDv
\,,&\textrm{with}\quad \eta \ge 0\,,
\end{array}
\fe
where $\vec{\kappa}$ is an arbitrary vector characterizing a
non-dissipative Magnus-type force orthogonal to the relative
velocity. 
Further substituting the conjugate momenta in the expression for the
hydrodynamic force densities (\ref{eq:DeffH}), we find their explicit
form 
\be
\vfH^\n\!\!=\!\!n_\n (\partial_t\!+\!\vv_\n\!\cdot\!\vnabla)\!\left( m
  \vv_\n\!+\!{2\entr\over n_\n}\vDv\!\right)\!+\! n_\n\!\vnabla \mu^\n\!+
\!2\entr \Dv_j \vnabla v_\n^j\,, \label{eq:NSvp1}
\fe
\begin{eqnarray}
\vfH^\c\!\!&=&\!\!n_\p (\partial_t\!+\!\vv_\c\!\cdot\!\vnabla)\!\left( m \vv_\c
\!-\!{2(\entr^{\e\n}\!+\!\entr^{\p\n})\over n_\p} \vDv\!\right)\!+\!n_\p
  \vnabla (\mu^\p\!+\!\mu^\e)\nonumber\\
& &\!\!\!- 2\entr \Dv_j \vnabla v_\c^j  - s(\partial_t +
  \vv_\c\cdot\vnabla)\left({2\entr^{\s\n}\over s}\vDv\right) +
  s\vnabla T\,. \label{eq:NSvp2}
\end{eqnarray}
We now make the simplifying assumption that we can neglect the
entrainment of entropy, i.e. we assume that all the entrainment
between the two fluids is due to the neutron-proton  and
neutron-electron contributions, so we set $\entr^{\s\n} = 0$, which
implies $\vec{\T} = 0$. Using (\ref{equpA}) we find
\begin{equation}
  \vp^\e + \vp^\p - \vp^\n = m 
  \left( 1 - \entrn_\n - \entrn_\c \right) \vDv \,,
\end{equation}
where we have defined the entrainment numbers  
\begin{equation}
\entrn_\n \equiv {2\entr\over\rho_\n}\,,\qaq
\entrn_\c \equiv {2\entr \over \rho_\c}\,.
\end{equation}
Putting all the pieces together, we obtain the momentum equations
(\ref{eq:NSfmutual}) and (\ref{eq:NSefHc}) in the form
\be
(\partial_t\!+\!\vv_\n\!\cdot\!\vnabla)\!\left(\vv_\n\!+\!\entrn_\n \vDv \right)
\!+\!\vnabla\left(\mut^\n\!+\!\Phi\right)\!+\!\entrn_\n \Dv_j \vnabla
v_\n^j\!=\!{1\over\rho_\n}\vcf_\mutual\,,\\ 
\fe
\bea
(\partial_t\!\!&+&\!\vv_\c\!\cdot\!\vnabla)\!\left(\vv_\c
\!-\!\entrn_\c \vDv\right)\!+\!\vnabla\left(\mut^\c\!+\!\Phi\right)
\!-\!\entrn_\c \Dv_j \vnabla v_\c^j\!+
\!{s\over\rho_\c}\vnabla T \nonumber\\
& &= - {1\over \rho_\c}\vcf_\mutual + \left(1-\entrn_\c -\entrn_\n\right) m
{\Rate_\n \over \rho_\c} \vDv\,.  \hspace{-1cm}
\fea
with the specific chemical potentials \mb{$\mut^\n\equiv \mu^\n/m$}
and \mb{$\mut^\c\equiv(\mu^\p+\mu^\e)/m$}.

\begin{acknowledgments}
  I would like to thank Brandon Carter and David Langlois for many
  valuable discussions about the relativistic variational principle
  and superfluids. I am also very grateful to Greg Comer and Nils
  Andersson for many helpful comments, and for interesting discussions
  about the superfluid neutron star model.
  I acknowledge support from the EU Programme 'Improving the Human 
  Research Potential and the Socio-Economic Knowledge Base' (Research
  Training Network Contract HPRN-CT-2000-00137).
\end{acknowledgments}

\appendix

\section{Evaluation of convective variations}
\label{sec:Variations}

We write the particle flowlines as 
\be
x^i = x^i (\va,t)\,,\label{equTrajectory}
\fe
where the ``particle coordinates'' $a^i$ are used to label individual
particles and can be taken, for example, to be their initial position,
i.e. 
\be
a^i = x^i(\va, 0)\,.
\fe
This introduces a time-dependent map (or ``pull-back'') between the
``material space'' $a^i$ and physical space $x^i$, and the associated
Jacobian matrix $\Jac$ is 
\be
{\Jac^i}_j \equiv \left.{\partial x^i \over \partial a^j}\right|_t\,.
\label{eq:DefJacobian}
\fe
We consider the variations of fluid variables induced by \emph{active}
infinitesimal spatial displacements $\xi^i(\vx,t)$ and temporal shifts
$\dtau(\vx,t)$ of the fluid particle flowlines (\ref{equTrajectory}),
namely 
\be
x'^i(\va,t') = x^i(\va,t) + \xi^i(\vx,t)\,,\qaq
t' = t + \dtau(\vx,t)\,. \label{equDisplacements1}
\fe
We note that the transformation (\ref{equDisplacements1}) not only
shifts flowlines in space, but also in time.
A physical quantity of the flow, $Q(\vx,t)$ say, is changed to  
$Q'(\vx',t')$, and we define the corresponding {\em Eulerian} and
{\em Lagrangian} variations as\footnote{Contrary to the Eulerian
  variation, the Lagrangian variation can be defined in different
  (non-equivalent) ways. The definition used here is based on
  comparing the quantity $Q$ in different points by parallel-transport. 
  Another common definition (e.g. see
  \cite{carter73:_elast_pertur,friedman78:_lagran}) consists in using
  the Lie-transported quantity instead. Both definitions are
equivalent for scalars but differ for vectors and higher order tensors.} 
\bea
\d Q\!\!&\equiv&\!Q'(\vx,t) - Q(\vx,t)\,, \label{equEulerianD}\\ 
\D Q\!\!&\equiv&\!Q'(\va,t')\!-\!Q(\va,t)\!=\!Q'(\vx',t')\!-\!Q(\vx,t)\,. \label{equLagrangianD} 
\fea
By expanding $\D Q$ to first order using the definition
(\ref{equDisplacements1}) of ${x^i}'$ and $t'$, we find the relation
\be
\D Q = \d Q + \xi^j\,\nabla_j Q(\vx,t) + \dtau \,\partial_t Q(\vx,t)\,.
\label{equEulLagr1}
\fe
Let us consider the induced (first order) variation of the
velocity \mb{$v^i\equiv\partial_t x^i(\va,t)$}, namely
\bea
v'^{i}(\va,t') &=& 
\partial_{t'} x'^{i}(\va,t') = \partial_{t'} x^i(\va,t) +  \partial_t \xi^i(\va,t) \nonumber\\
&=& \partial_t x^i(\va,t)\, \left.{\partial t \over \partial t'}\right|_\va
      + \partial_t \xi^i(\va,t)\nonumber\\ 
&=& v^i(\va,t) - v^i\, \partial_t \dtau(\va,t) + \partial_t \xi^i(\va,t)\,,
\fea
which by (\ref{equLagrangianD}) corresponds to the following
Lagrangian variation of the velocity:
\be
\D v^i = \left[ \partial_t \xi^i + v^l \nabla_l \xi^i \right] 
- \left[v^i \partial_t \dtau + v^i v^l \nabla_l \dtau  \right]\,,
\fe
and with (\ref{equEulLagr1}) the Eulerian variation is found as
\be
\d v^i = \left[\partial_t \xi^i + v^l \nabla_l \xi^i - \xi^l \nabla_l v^i \right]
- \left[\partial_t\left(v^i \dtau \right) + v^i v^l \nabla_l \dtau \right]\,.
\fe
From the conservation of mass one can derive an expression for the
particle density $n$ in terms of the Jacobian (\ref{eq:DefJacobian}),
namely
\be
n (\vx,t) = {n_0(\va) \over \det \Jac }\,,
\label{eq:Density}
\fe
where $n_0(\va)=n(\va,0)$ is the initial density at $t=0$. 
Using (\ref{eq:DefJacobian}), the change of the Jacobian matrix $\Jac$ 
induced by the flowline variation (\ref{equDisplacements1}) can be
found as 
\bea
{\Jac'^i}_j(\va,t') &=& {\partial x'^i(\va,t') \over \partial a^j } = 
\left.{\partial x^i(\va,t) \over \partial a^j }\right|_{t'} + 
{\partial \xi^i \over \partial a^j } \nonumber\\
&=& {\partial x^i(\va,t) \over \partial a^j} + {\partial x^i(\va,t)
  \over \partial t}\, \left.{\partial t \over \partial a^j}\right|_{t'}
 + {\partial \xi^i \over \partial a^j }\nonumber\\
&=& {\Jac^i}_j (\va,t) - v^i {\partial \dtau \over \partial a^j}
 + {\partial \xi^i \over \partial a^j }\,,
\fea
with the resulting Lagrangian variation (\ref{equLagrangianD}) expressible as
\be
\D {\Jac^i}_j = {\Jac^l}_j \left( \nabla_l \xi^i - v^i \nabla_l \dtau \right)\,.
\fe
The derivative of a determinant $\det A$ with respect to a matrix
element $A_{i j}$ is given by  
\be
{\partial \det A \over \partial A_{i j}} = \det(A)\,\left(A^{-1}\right)^{i j}\,,
\label{eq:dDet}
\fe
and therefore we can write the Lagrangian variation of the Jacobian
determinant as
\be
\D \left(\det \Jac\right) = \det(\Jac)\, {\left(\Jac^{-1}\right)^j}_i \, \D {\Jac^i}_j\,.
\fe
The flowline variation (\ref{equDisplacements1}) therefore induces the
Lagrangian change of the Jacobian
\be
{\D\left( \det J\right) \over \det J}= \nabla_l \xi^l - v^l \nabla_l \dtau\,.
\fe
Using (\ref{eq:Density}), the induced density variation is therefore
found as 
\be
\D n = - n \nabla_l \xi^l + n v^l \nabla_l \dtau \,,
\fe
and with (\ref{equEulLagr1}) the corresponding Eulerian expression is
found as
\be
\d n = - \nabla_l\left( n \xi^l \right) + \left[n v^l \nabla_l \dtau -
\dtau \partial_t n \right]\,.
\label{eq:dDensity}
\fe
By combining the results for velocity and density variations we find
the variations of the current \mb{$n^i = n v^i$} as 
\be
\D n^i = \left[ n\,\partial_t \xi^i(\vx,t) + n^l\nabla_l \xi^i -  
n^i \nabla_l \xi^l \right] -  n^i \, \partial_t \dtau\,,
\fe
\be
\d n^i\!=\!\left[ n\partial_t \xi^i(\vx,t)\!+\!n^l\nabla_l \xi^i\!- 
\!\nabla_l\left(n^i\xi^l\right)\right]\!-\!\partial_t \left(n^i \dtau \right)\,.
\label{eq:dCurrent}
\fe

\section{Noether identities of the variational principle}
\label{sec:VariationalTij}

In addition to the flowline variations considered so far, we will now
also allow for \emph{metric variations} $\d g_{i j}$. Although
we only consider Newtonian physics here, there is a-priori no reason
to restrict ourselves to flat space. Most importantly, however,
including metric variations allows us to obtain the form of the stress
tensor $T_\H^{i j}$ and the associated momentum conservation
(\ref{eq:MomCons}) directly from the variational principle as a Noether
identity, as opposed to constructing it from the equations of motion
as we have done in Sec.~\ref{sec:TotalConservationLaws}.
Therefore we extend the variation (\ref{equVarL}) of the Lagrangian
to
\be
\d \LagrH = \csum \po^\X\, \d n_\X + \csum \vp^\X\cdot\d\vn^\X +
{\partial \LagrH \over \partial g_{i j}}\, \d g_{i j}\,.
\label{eq:dMetricLagr}
\fe
Next consider the density change $\d n^\X$ induced by a metric
variation $\d g_{i j}$ at constant flowlines, i.e. constant ${\Jac^i}_j$.
First we note that we can express the Jacobian as
\be
\det \Jac  = \eps_{i j k}\, {\Jac^i}_1 \, {\Jac^j}_2\, {\Jac^k}_3\,,
\fe
and using (\ref{eq:dDet}) the variation of the volume form 
\mb{$\eps_{i j k}=\sqrt{g}\left[i j k\right]$} induced by metric
changes is expressible as
\be
\d \eps_{i j k} = {1\over 2}\eps_{i j k}\, g^{l m}\d g_{l m}\,.
\fe
Therefore we have
\be
\left.{\partial \det \Jac \over \partial g_{i j}}\right|_{\Jac} =
{1\over 2} \det(\Jac) \, g^{i j}\,,
\fe
and using  (\ref{eq:Density}) and (\ref{eq:dDensity}) we can write the
variation of the density induced by spatial displacements $\vxi$ and
metric variations $\d g_{i j}$ as 
\bea
\d n &=& - \nabla_l \left( n \xi^l \right) - {1\over 2} n g^{i j}\, \d g_{i j}\,.
\label{eq:dMetricDens} \\
\D n &=& - n \nabla_l \xi^l - {1\over 2} n g^{i j}\, \d g_{i j}\,,
\label{eq:DMetricDens}
\fea
where we have used the fact that with our definition of the Lagrangian
variation (\ref{equEulLagr1}) we have
\be
\D g_{i j} = \d g_{i j} + \xi^l\nabla_l g_{i j} = \d g_{i j}\,,
\fe
as the metric is by definition constant under parallel transport.
A metric change with fixed flowlines does not change the local velocity
$v^i$, therefore the current variation can be written using
(\ref{eq:dMetricDens}) and (\ref{eq:dCurrent}) as
\be
\d n^i\!=\!\left[ n\,\partial_t \xi^i(\vx,t)\!+\!n^l\nabla_l \xi^i\!-\!
\nabla_l\left(n^i\xi^l\right)\right] 
\!-\!{1\over 2} n^i g^{l j} \d g_{l j}\,,
\label{eq:dMetricCurrent}
\fe
\be
\D n^i = \left[ n\,\partial_t \xi^i(\vx,t) + n^l\nabla_l \xi^i -  
n^i \nabla_l \xi^l \right] - {1\over 2} n^i \,g^{l j} \d g_{l j}\,.
\label{eq:DMetricCurrent}
\fe
When allowing for metric variations it is convenient
(e.g. see \cite{carter98:_relat_supercond_superfl}) to introduce the
``diamond variation'' $\Diamond\LagrH$ as
\be
\Diamond \LagrH \equiv {1\over \sqrt{g}} \d\left( \sqrt{g}\,\LagrH\right)
= \d\LagrH + {1\over2}\LagrH \, g^{i j}\d g_{i j}\,,
\fe
such that the variation of the action (\ref{eq:DefAct}) can now be written
as (noting that \mb{$d V = \sqrt{g}\,d^3 x$}):
\be
\d\Act = \int \Diamond\LagrH \, d V\, d t\,.
\fe
Substituting (\ref{eq:dMetricLagr}), (\ref{eq:dMetricDens}) and
(\ref{eq:dMetricCurrent}) and integrating by parts, 
\mb{$\Diamond \LagrH$} can be cast in the form
\be
\Diamond \LagrH = - \csum f^\X_i \, \xi_\X^i + {1\over2} T_\H^{i
  j}\,\d g_{i j} + \nabla_l R^l + \partial_t R\,,
\label{eq:DiamondLagr}
\fe
where the canonical forces $\vf_\X$ have the explicit expression
(\ref{equfA}) and we defined the tensor $T_\H^{i j}$ as
\be
T_\H^{i j} \equiv 2 {\partial \LagrH \over \partial g_{i j}} +
\Press\,g^{i j}\,,
\label{eq:NoetherTij}
\fe
using our earlier definition (\ref{equDefPress}) of the generalized
pressure $\Press$.

Now consider a common displacement $\vxi$ of the \emph{whole} system
including the background metric, which induces a metric change
\be
\d g_{i j} = - 2 \nabla_{(i} \xi_{j)}\,,
\label{eq:MetricVar}
\fe
where $(i j)$ indicates symmetric averaging, i.e. \mb{$2 A_{(i}B_{j)}=A_i B_j + A_j B_i$}.
The corresponding Lagrangian variations
(\ref{eq:DMetricCurrent}) and (\ref{eq:DMetricDens}) are found as
\bea
\D n_\X &=& 0\,,\\
\D n_\X^i &=& n_\X \left( \partial_t \xi^i + v_\X^l \nabla_l \xi^i \right)\,.
\fea
Substituting this into (\ref{eq:dMetricLagr}), the induced $\D \LagrH$ is
\be
\D \LagrH = \left( \csum n_\X^i p^{\X\,j} - 2 {\partial \LagrH \over
    \partial g_{i j} } \right)\, \nabla_i \xi_j + J_\H^i\, \partial_t\xi_i\,,
\label{eq:DMetricLagr}
\fe
where we have used the definition (\ref{eq:Tmunu}) of the momentum
density $\vJ_\H$. It is well known that contrary to the fully covariant
Lagrangian for relativistic hydrodynamics
(e.g. \cite{carter89:_covar_theor_conduc}), the Newtonian Lagrangian
is not strictly Galilean invariant under boosts. This is due to the
velocity dependence of the kinetic energy, as can be seen in the
explicit form (\ref{eq:DefLagrH}).
We can therefore only demand strict invariance, i.e. 
\mb{$\D \LagrH =0$},  for time-independent displacements, namely
\mb{$\partial_t\vxi=0$}, which leads to the Noether identity 
\be
{\partial \LagrH \over \partial g_{i j}} = {1\over2} \csum n_\X^i
p^{\X \, j}\,.
\fe
The left-hand side is manifestly symmetric in $i$ and $j$, therefore
we see that
\be
\csum  n_\X^i p^{\X j} = \csum n_\X^j p^{\X i}\,,
\label{eq:NoetherSymmetry}
\fe
and we can now write the (symmetric) stress tensor (\ref{eq:NoetherTij})
explicitly as
\be
{T_\H^i}_j  = \csum n_\X^i \, p^\X_j + \Press\, {g^i}_j \,.
\fe
This tensor is identical to the expression (\ref{eq:Tmunu}) found
earlier by construction from the equations of motion. It remains to be
shown however, how the momentum conservation law (\ref{eq:MomCons}) is
directly obtainable as a Noether identity from the variational principle. 
Using (\ref{eq:DMetricLagr}), (\ref{equEulLagr1}) and
(\ref{eq:DiamondLagr}) we can explicitly express the diamond variation as
\be
\Diamond \LagrH = -(\partial_t J^j)\, \xi_j - \nabla_l(\LagrH\, \xi^l) 
+ \partial_t(J_\H^l\, \xi_l)\,,
\fe
which has to be identical to the expression (\ref{eq:DiamondLagr}) for
a common displacement $\vxi$ of the whole system, which after
some partial integrations takes the form
\be
\Diamond \LagrH = \left( -\csum f^{\X\,j} + \nabla_l T_\H^{l j}\right)
\, \xi_j + \nabla_l(...)^l + \partial_t(...)\,.
\fe
The requirement that the previous two expressions have to be identical
(up to divergences and time derivatives) leads to the Noether identity
\be
\partial_t J_\H^i + \nabla_j T_\H^{i j} = f_\ext^i\,,
\fe
which is the momentum conservation law (\ref{eq:MomCons}).

\section{Galilean invariance of $\E$}
\label{sec:GalileanE}

In this section we show that requiring the conjugate momenta $\po^\X$
and $\vp^\X$ of (\ref{eq:HydroMomenta}) to transform as
(\ref{eq:pX_boost}) under Galilean boosts (\ref{eq:GalileanBoost})
implies that the internal energy $\E$ has to be Galilean invariant.
We assume that $\E(n_\X, \vn_\X)$ transforms into $\E'(n_\X, \vn_\X')$
under a Galilean boost, where 
\begin{equation}
\vn_\X' = \vn_\X + n_\X \vV\,.  
\label{eq:A1}
\end{equation}
Therefore the conjugate momenta (\ref{eq:HydroMomenta}) in the frame
moving with speed $-\vV$ are of the form 
\begin{eqnarray}
  -{\po^\X}'\!\!&=&\!{1\over 2}m^\X \vv_\X^2\!+\!m^\X\vv_\X\!\cdot\!\vV
\!+\!{1\over2} m^\X \vV^2\!+\!{\partial \E' \over \partial n_\X}\,,\\
{\vp^\X}{}'\!\! &=&\! m^\X \vv_\X + m^\X \vV - {\partial \E' \over \partial \vn_\X'}\,,
\end{eqnarray}
Using (\ref{eq:HydroMomenta}) to eliminate all terms containing
$\vv_\X$, we arrive at  
\be
  -{\po^\X}'\!\!=\!-\po^\X\!+\!\vV\!\cdot\!\vp^\X\!+\!{1\over2}m^\X \vV^2
  \!+\!\left[{\partial \E' \over \partial n_\X}
    \!-\!{\partial \E \over n_\X}\!+\!\vV\!\cdot\!{\partial \E \over
   \vn_\X}\right]\,, \label{eq:A3}
\fe
\be
{\vp^\X}{}'\!\!=\!\vp + m^\X \vV + \left[{\partial \E \over \partial
   \vn_\X} - {\partial \E' \over \partial \vn_\X'} \right]\,. \label{eq:A4}
\fe
By comparing with the required transformation properties
(\ref{eq:pX_boost}) we see that a necessary and sufficient condition
for this is the vanishing of the terms in brackets in (\ref{eq:A3})
and (\ref{eq:A4}). 
We can rewrite the partial derivatives of the energy function as follows
\begin{equation}
  {\partial \E' \over \partial \vn_\X'} = {\partial \E' \over \partial
  \vn_\X}\cdot \left.{\partial \vn_\X \over \partial \vn_\X'}\right|_{n_\X} = 
{\partial \E' \over \partial \vn_\X}\,,
\end{equation}
and
\begin{equation}
  \left.{\partial \E' \over \partial n_\X}\right|_{\vn_\X'}\!\!=\!
  \left.{\partial \E' \over \partial n_\X}\right|_{\vn_\X}\!+\!
  {\partial \E' \over \partial \vn_\X}\!\cdot\!
  \left.{\partial \vn_\X \over \partial n_\X}\right|_{\vn_\X'}\!=\!
  \left.{\partial \E' \over \partial n_\X}\right|_{\vn_\X}\!-\!
  \vV\!\cdot\!  {\partial \E' \over \partial \vn_\X}\,.
\end{equation}
Inserting these identities into (\ref{eq:A3}) and (\ref{eq:A4}), the
invariance requirement can be expressed as
\begin{equation}
  \left.{\partial \E \over \partial n_\X}\right|_{\vn_\X} = 
  \left.{\partial \E' \over \partial n_\X}\right|_{\vn_\X}\,,\qaq
  {\partial \E \over \partial \vn_\X} = 
  {\partial \E' \over \partial \vn_\X}\,,\quad\textrm{for all}\;\X\,,
\end{equation}
therefore $\E'$ can only differ from $\E$ by a constant, which is
unimportant because the absolute value of the energy scale is
arbitrary. This shows that energy function $\E$ has to be Galilean
invariant under the above assumptions.

\section{Newtonian limit of the relativistic Lagrangian}
\label{sec:NewtLimit}

As shown in the relativistically covariant framework by Carter
\cite{carter89:_covar_theor_conduc}, the equations of motion for 
conducting multi-constituent fluids can be derived from a covariant
Lagrangian density of the form 
\begin{equation}
\Lagr_\cov = - \rho c^2\,,  
\label{eq:LagrCov}
\end{equation}
where the scalar $\rho$ is now the total mass-energy density of the system.
For simplicity we consider here a two-fluid system, as generalizations
to more fluids are straightforward while making the notation more cumbersome.
The two fluids, $\A$ and $\B$ say, are described by the two 4-current
densities  $n_\A^\mu$, $n_\B^\mu$, and therefore the scalar
\mb{$\Lagr_\cov(n_\A^\mu, n_\B^\mu)$} can only depend on the three
independent scalar combinations of these two currents, for example
$$
  n_\A = {1\over c} \sqrt{- g_{\mu\nu}n_\A^\mu n_\A^\nu}\,,\quad
  n_\B = {1\over c} \sqrt{- g_{\mu\nu}n_\B^\mu n_\B^\nu}\,,
$$
and
\be
  x = {1\over c} \sqrt{- g_{\mu\nu}n_\A^\mu n_\B^\nu}\,,
\label{eq:CovnX}
\fe
and so generally \mb{$\Lagr_\cov = \Lagr_\cov(n_\A, n_\B, x)$}. 
Instead of $x$ we can equivalently choose as a third independent
quantity the combination  
\begin{equation}
  {\Dv^2 \over c^2} \equiv 1 - \left({n_\A n_\B \over x^2}\right)^2 \,.
\label{eq:CovRelVel}
\end{equation}
We are interested here only in the purely hydrodynamic content of this
framework, so we  assume a flat space-time, i.e. a metric of the form
\begin{equation}
  d s^2 = g_{\mu\nu} d x^\mu d x^\nu = -c^2 \,d t^2 + d\vx^2\,,
  \label{eq:FlatMetric}
\end{equation}
with the time-coordinate $x^0=t$ and so $g_{00}=-c^2$.
When taking the Newtonian limit as \mb{$c\rightarrow \infty$}, the
metric becomes singular. The reason for this singular limit
obviously lies in the fact that a locally Lorentzian theory reduces to
a Galilean invariant theory, therefore the Lorentz invariance has to
be broken in the limit. 
As the non-invertible metric no longer fully determines the
space-time, we now have to \emph{choose}\footnote{See
  \cite{carter94:_canon_formul_newton_superfl} for a more detailed
  discussion of this limit and how to construct a fully space-time
  covariant Newtonian framework.}
a preferred time coordinate, $t$ say, in which to take the limit and
which will reduce to the Newtonian absolute time. 

The relation between the scalar rest-frame particle densities
$n_\X$ and the densities $n_\X^0$ in the preferred-time frame can
be expressed from (\ref{eq:CovnX}) and (\ref{eq:FlatMetric}):
\begin{equation}
  n_\X\!=\!{1\over c}\sqrt{c^2 (n_\X^0)^2\! -\! \vn_\X^2}\!=\!
  n_\X^0\!\left[ 1\! -\! {1\over 2} \left({\vv_\X \over c} \right)^2
  \right]\!+\!\O\left(c^{-4} \right)\,, 
\label{eq:FrameDens}
\end{equation}
where \mb{$(\vn_\X)^i = n_\X^i$} is the spatial part of the 4-current
$n_\X^\mu$ in the preferred time frame, and the relation to the
Newtonian 3-velocity $\vv_\X$ is simply \mb{$\vn_\X = n_\X^0\, \vv_\X$}.
We see from this equation that if we choose the densities $n_\X^0$ to
represent the Newtonian particle number densities independent of $c$,
then in the limit we find
\begin{equation}
  \lim_{c\rightarrow \infty} n_\X = n_\X^0\,.
\label{eq:DensLimit}
\end{equation}
We further note that the quantity $\Dv$ introduced in
(\ref{eq:CovRelVel}) reduces to the relative velocity in the Newtonian 
limit, namely
\begin{equation}
  \Dv^2 = \left(\vv_\A - \vv_\B \right)^2 + \O\left( (v/c)^2 \right) \,.
\end{equation}
We now turn to the covariant Lagrangian $\Lagr_\cov$ of
(\ref{eq:LagrCov}) which we can quite generally be written as
\begin{equation}
  \Lagr_\cov = - (n_\A m_\A + n_\B m_\B)\, c^2 - \E(n_\A,n_\B,\Dv^2) \,,
\end{equation}
where the first term represents the rest-mass energy in the fluid
frame, while $\E$ contains the ``equation of state'', i.e. the
internal-energy function of the fluid.
When we write this in the preferred time-frame using
(\ref{eq:FrameDens}), we obtain
\bea
\Lagr_\cov\!\!&=&\!-(n_\A^0 m_\A\!+\!n_\B^0 m_\B) c^2 
\!+\!{1\over2} m_\A\, n_\A^0 \vv_\A^2\!+\!{1\over 2} m_\B n_\B^0 \vv_\B^2\nonumber\\
&&- \E(n^0_\A,n^0_\B,\Dv^2) + \O\left( (v/c)^2 \right)\,.
\fea
We see that this Lagrangian obviously diverges in the Newtonian limit
$c\rightarrow\infty$ due to the rest-mass energies $n_\X^0\,m_\X\,c^2$.
Before we can take this limit, we therefore have to renormalize the
Lagrangian density by subtracting a finite counter-term that will make
the limit finite.
The most natural choice is obviously to subtract the mass-energy in
the preferred-time frame that will determine the Newtonian absolute
time. We  therefore define the renormalized Lagrangian density
$\Lagr_\ren$ as  
\begin{equation}
  \Lagr_\ren \equiv \Lagr_\cov + (n_\A^0 m_\A + n_\B^0 m_\B) \,c^2\,.
\end{equation}
In  $\Lagr_\ren$ we have  explicitly broken Lorentz invariance by
choosing a preferred time frame, and when taking the Newtonian limit
we obtain the finite Lagrangian
\begin{equation}
\lim_{c\rightarrow\infty} \Lagr_\ren = m_\A\, {\vn_\A^2\over 2\, n_\A} + 
m_\B \,{\vn_\B^2 \over 2\, n_\B} - \E(n_\A, n_\B, \Dv^2)\,,
\end{equation}
which corresponds exactly to the Newtonian hydrodynamic Lagrangian
$\Lagr_\H$ of (\ref{equLH}). 

\bibliography{biblio}

\end{document}